%% file: bare_jrnl.tex
\PassOptionsToPackage{colorlinks=true}{hyperref}
\documentclass[journal, USenglish]{IEEEtran}

\input{TeX/Preamble.tex}

\makenomenclature

\begin{document}
	\input{TeX/glossary.tex}
	\input{TeX/nomenclature.tex}
	\title{Optimal Grid Layouts for Hybrid Offshore Assets in the North Sea under Different Market Designs}

	\author{\IEEEauthorblockN{Stephen Hardy$^{1}$
	\orcidlink{0000-0001-8893-8682}, Andreas Themelis$^{2}$ \orcidlink{0000-0002-6044-0169}, Kaoru Yamamoto$^{2}$~\IEEEmembership{Member,~IEEE,} \orcidlink{0000-0003-1888-3529},
			Hakan~Ergun$^{1}$~\IEEEmembership{Senior Member,~IEEE,} \orcidlink{0000-0001-5171-1986} and~Dirk~Van Hertem$^{1}$~\IEEEmembership{Senior Member,~IEEE,} \orcidlink{0000-0001-5461-8891}}%
		\texorpdfstring{\\}{ }%
		\IEEEauthorblockA{$^1$KU Leuven/EnergyVille, Leuven/Genk, Belgium\quad $^2$Kyushu University, Fukuoka, Japan}
		\thanks{This project has received funding from the CORDOBA project funded by Flanders Innovation and Entrepreneurship (VLAIO) in the framework of the spearhead cluster for blue growth in Flanders (Blue Cluster) – Grant number HBC.2020.2722, and from the Japan Society for the Promotion of Science (JSPS) KAKENHI grants no. JP19H02161, JP20K14766 and JP21K17710. A special thanks to Michiel Kenis and Prof. Jef Berten for their input regarding energy market modelling.}%
	}

	\markboth{}%
	{Hardy \MakeLowercase{\textit{et al.}}: Optimal Grid Layouts for Hybrid Offshore Assets in the North Sea under Different Market Designs.}

	\maketitle
	\begin{abstract}
		\input{TeX/abstract.tex}
	\end{abstract}


	\printnomenclature

	\IEEEpeerreviewmaketitle

	\section{Introduction}
		\input{TeX/introduction.tex}

	\section{Nodal versus zonal markets}\label{sect: pivotal_supplier}
		\input{TeX/nodalVSzonal.tex}

	\section{Planning model}
		\input{TeX/planning.tex}

	\section{Test Case}\label{sect:test_cases}
		\input{TeX/testCase.tex}

	\vspace{-3mm}
	\section{Results}\label{sect:results}
		\input{TeX/results.tex}

	\vspace{-3mm}
	\section{Conclusion}\label{sect:conclusions}
		\input{TeX/conclusion.tex}

	\ifCLASSOPTIONcaptionsoff
	\newpage
	\fi

	\vspace{-2.5mm}
	\bibliographystyle{IEEEtran}
	\bibliography{references}
	\newpage

	\section{Appendix}
		\input{TeX/appendix.tex}

\end{document}

%% file: TeX/Preamble.tex
\usepackage{cite}
\usepackage[dvipsnames]{xcolor}
\usepackage{amsmath}
\usepackage[nopostdot,style=super,nonumberlist,toc]{glossaries}
\usepackage{bm}
\usepackage{verbatim}
\usepackage{makecell}
\DeclareMathAlphabet{\mathpzc}{OT1}{pzc}{m}{it}
\usepackage{tikz}
\usetikzlibrary{shapes, arrows, arrows.meta, positioning,calc}
\usepackage{colortbl}
\definecolor{light-gray}{gray}{0.90}
\usepackage{float}
\usepackage{babel}
\usepackage{caption}
\usepackage{orcidlink}
\usepackage{pgfplots}
\pgfplotsset{compat=1.16}
\usepackage{xurl}
\usepackage{etoolbox}
\usepackage{multicol}
\usepackage{nomencl}
\usepackage{mathtools}

\definecolor{findOptimalPartition}{HTML}{D7191C}
\definecolor{storeClusterComponent}{HTML}{FDAE61}
\definecolor{dbscan}{HTML}{ABDDA4}
\definecolor{constructCluster}{HTML}{2B83BA}
\definecolor{rosso}{RGB}{232,76,61}
\definecolor{giallo}{RGB}{64,64,64}
\definecolor{blu}{RGB}{217,217,217}
\definecolor{verde}{RGB}{242,242,242}
\definecolor{viola}{RGB}{153,0,153}
\definecolor{Gray}{gray}{0.85}	
\definecolor{aliceblue}{rgb}{0.94, 0.97, 1.0}
\definecolor{LightCyan}{rgb}{0.88,1,1}	
\definecolor{anti-flashwhite}{rgb}{0.95, 0.95, 0.96}
\newcolumntype{d}{>{\columncolor{Gray}}m{6mm}}
\newcolumntype{e}{>{\columncolor{LightCyan}}m{6mm}}
\newcolumntype{f}{>{\columncolor{aliceblue}}m{6mm}}
\newcolumntype{a}{>{\columncolor{anti-flashwhite}}m{6mm}}
\newcolumntype{D}{>{\columncolor{Gray}}m{6mm}}
\newcolumntype{E}{>{\columncolor{LightCyan}}m{6mm}}
\newcolumntype{F}{>{\columncolor{aliceblue}}m{7mm}}
\newcolumntype{A}{>{\columncolor{anti-flashwhite}}m{6mm}}
\newcolumntype{z}{>{\columncolor{Gray}}m{5.5mm}}
\newcolumntype{x}{>{\columncolor{LightCyan}}m{5.5mm}}
\newcolumntype{y}{>{\columncolor{aliceblue}}m{5.5mm}}
\newcolumntype{v}{>{\columncolor{anti-flashwhite}}m{5.5mm}}
\newcommand{\red}{\color{red}}

\newcommand{\Z}{\mathcal{Z}}	

\let\EA\expandafter

	\let\OLDwidetilde\widetilde
	\let\OLDtilde\tilde
	\let\OLDbar\bar
	\let\OLDoverline\overline

	\renewcommand{\widetilde}[1]{\OLDwidetilde #1{\red<!>}}
	\renewcommand{\tilde}[1]{\OLDtilde #1{\red<!>}}
	\renewcommand{\bar}[1]{\OLDbar #1{\red<!>}}
	\renewcommand{\overline}[1]{\OLDoverline #1{\red<!>}}

	\newcommand{\Set}{\mathcal S}
	\newcommand{\A}{\mathcal A}
	\newcommand{\Exist}[1]{\EA\OLDbar #1}
	\newcommand{\Cand}[1]{\EA\OLDwidetilde #1}

	\newcommand{\ac}{{\text{\sc ac}}}
	\newcommand{\dc}{{\text{\sc dc}}}
	\newcommand{\N}{\mathcal{N}}	
	\newcommand{\E}{\mathcal{E}}	
	\newcommand{\Eacdc}{\E^{\ac\text-\dc}}	

	\newcommand{\linebetween}[2]{{\scriptscriptstyle\{}#1#2{\scriptscriptstyle\}}}
	\newcommand{\mnm}{\linebetween{m}{n}}
	\newcommand{\efe}{\linebetween{e}{f}}

	\newcommand{\Ss}{\Set_{\rm s}}					
	\newcommand{\St}{\Set_{\rm t}}					
	\newcommand{\Sy}{\Set_{\rm y}}					

	\newcommand{\Sj}{\Set_{\rm j}}					
	\newcommand{\Sjn}{\Set_{{\rm j}:n}}					

	\newcommand{\Sg}{\Set_{\rm g}}					
	\newcommand{\Sgn}{\Set_{{\rm g}:n}}					

	\newcommand{\Su}{\Set_{\rm u}}					
	\newcommand{\Sun}{\Set_{{\rm u}:n}}					

	\newcommand{\Sl}{\Set_\ell}						
	\newcommand{\Slmnm}{\Set_{\ell:\mnm}}				
	\newcommand{\Slefe}{\Set_{\ell:\efe}}				

%% file: TeX/glossary.tex
\newacronym{MILP}{MILP}{Mixed Integer Linear Program}
\newacronym{MIP}{MIP}{Mixed Integer Program}
\newacronym{EEA}{EEA}{Exclusive Economic Area}
\newacronym{OWPP}{OWPP}{Offshore Wind Power Plant}
\newacronym[firstplural=Points of Common Coupling]{PCC}{PCC}{Point of Common Coupling}
\newacronym{FOWIND}{FOWIND}{Facilitating Offshore Wind in India}
\newacronym{GEC}{GEC}{Green Energy Corridor}
\newacronym{NIWE}{NIWE}{National Institute of Wind Energy}
\newacronym{DTU}{DTU}{Denmark Technical University}
\newacronym{EENS}{EENS}{Expected Energy Not Served}
\newacronym{EENT}{EENT}{Expected Energy Not Transmitted}
\newacronym{OSS}{OSS}{Offshore Substation}
\newacronym{NGZ}{NGZ}{No-Go Zone}
\newacronym{NGR}{NGR}{No-Go Region}
\newacronym{GPS}{GPS}{Global Positioning System}
\newacronym{NPV}{NPV}{Net Present Value}
\newacronym{RES}{RES}{Renewable Energy Source}
\newacronym{lsb}{lsb}{Least Significant Bit} 
\newacronym{nsb}{nsb}{Next Significant Bit}
\newacronym{msb}{msb}{Most Significant Bit}
\newacronym{TNEP}{TNEP}{Transmission Network Expansion Planning}
\newacronym{TSO}{TSO}{Transmission System Operator}
\newacronym{MOG}{MOG}{Modular Offshore Grid}
\newacronym{CAPEX}{CAPEX}{Capital Expenditures}
\newacronym{OFGEM}{OFGEM}{Office of Gas and Electricity Markets}
\newacronym{OFTO}{OFTO}{Offshore Transmission Owner}
\newacronym{CSI}{CSI}{Cross Sub-Station Integration}
\newacronym{MV}{MV}{Medium Voltage}
\newacronym{HV}{HV}{High Voltage}
\newacronym{NSWPH}{NSWPH}{North Sea Wind Power Hub}
\newacronym{COPT}{COPT}{Capacity Outage Probability Table}
\newacronym{CM}{CM}{Corrective Maintenance}
\newacronym{CE}{CE}{Constrained Energy}
\newacronym{HVAC}{HVAC}{High Voltage Alternating Current}
\newacronym{HVDC}{HVDC}{High Voltage Direct Current}
\newacronym{OWTOP}{OWTOP}{Offshore Wind Topology Optimization Problem}
\newacronym{ADMM}{ADMM}{Alternating Direction Method of Multipliers}
\newacronym{TYNDP}{TYNDP}{Ten Year Network Development Plan}
\newacronym{GATE}{GATE}{Generation and Transmission Expansion}
\newacronym{NT}{NT}{National Trends}
\newacronym{DG}{DG}{Distributed Generation}
\newacronym{GA}{GA}{Global Ambition}
\newacronym[firstplural=National Energy Climate Policies (NECPs)]{NECP}{NECP}{National Energy Climate Policy}
\newacronym{EC}{EC}{European Commission}
\newacronym{ENTSO-E}{ENTSO-E}{European Network of Transmission System Operators for Electricity}
\newacronym{HOA}{HOA}{Hybrid Offshore Asset}
\newacronym{HMD}{HMD}{Home Market Design}
\newacronym{nOBZ}{nOBZ}{nodal Offshore Bidding Zone}
\newacronym{zOBZ}{zOBZ}{zonal Offshore Bidding Zone}
\newacronym{EEZ}{EEZ}{Exclusive Economic Zone}
\newacronym{GCS}{GCS}{Gross Consumer Surplus}
\newacronym{MINLP}{MINLP}{Mixed Integer Non-Linear Program}
\newacronym{OBZ}{OBZ}{Offshore Bidding Zone}
\newacronym{RDCC}{RDCC}{Regulatory re-Dispatch with Cost Compensation}
\newacronym{DSR}{DSR}{Demand Side Response}
\newacronym{VOLL}{VOLL}{Value Of Lost Load}
\newacronym[firstplural=Net Transfer Capacities (NTCs)]{NTC}{NTC}{Net Transfer Capacity}

%% file: TeX/nomenclature.tex
	\nomenclature{$\Ss$}{Set of scenarios}
	\nomenclature{$\St$}{Set of hours}
	\nomenclature{$\Sy$}{Set of years}

	\nomenclature{$\Sj$}{Set of storage devices}

	\nomenclature{$\Sg$}{Set of all generators}					
	\nomenclature{$\Exist \Sg$}{Set of existing generators}
	\nomenclature{$\Cand \Sg$}{Set of candidate generators}

	\nomenclature{$\Su$}{Set of demands}

	\nomenclature{$\Sl$}{Set of all transmission lines}
	\nomenclature{$\Exist \Sl$}{Set of existing transmission lines}
	\nomenclature{$\Cand \Sl$}{Set of candidate transmission lines}
	
	\nomenclature{$f^{\text{\sc y}}$}{NPV scalar for yearly revenues}
	\nomenclature{$f^{\text{\sc h}}$}{NPV scalar for hourly revenues}
	
	\nomenclature{$\N$}{Set of all nodes}	
	\nomenclature{$\N^\ac$}{Set of all AC nodes}	
	\nomenclature{$\N^\dc$}{Set of all DC nodes}	
	\nomenclature{$\E$}{Set of all (directed) edges}	
	\nomenclature{$\E^\ac$}{Set of AC network edges}	
	\nomenclature{$\E^\dc$}{Set of DC network edges}	
	\nomenclature{$\E^{\rm br}$}{Set of intra-zonal edges}	
	\nomenclature{$\E^{\rm te}$}{Set of inter-zonal edges}	
	\nomenclature{$\Eacdc$}{Set of all edges between AC and DC networks}	
\nomenclature{$\lambda$}{Market clearing price}
\nomenclature{$C^{\rm g}$}{Generator bid price}
\nomenclature{$C^{\rm u}$}{Consumer bid price}

\nomenclature{$P^{\rm g}$}{Instantaneous generator power}
\nomenclature{$P^{\rm u}$}{Instantaneous demand power}
\nomenclature{$P^{\rm j,abs}$}{Instantaneous storage charging power}
\nomenclature{$P^{\rm j,inj}$}{Instantaneous storage discharging power}
\nomenclature{$P^{\ell}$}{Instantaneous Transmission line power}
\nomenclature{$P^{\zeta,\ac}$}{Instantaneous AC side converter power}
\nomenclature{$P^{\zeta,\dc}$}{Instantaneous DC side converter power}
\nomenclature{$P^{\zeta,\text{\rm loss}}$}{Instantaneous converter power losses}
\nomenclature{$P^{\rm g,max}$}{Maximum generator power}
\nomenclature{$P^{\rm u,max}$}{Maximum demand power}
\nomenclature{$P^{\ell, \rm max}$}{Maximum  transmission line power}
\nomenclature{$P^{\zeta, \rm max}$}{Maximum converter power}
\nomenclature{$\widehat{P^{\zeta,\rm max}}$}{Maximum expansion capacity of candidate converters}
\nomenclature{$\widehat{P^{\rm g, max}}$}{Maximum expansion capacity of candidate generators}

\nomenclature{$\delta I^{{\rm g}}$}{Generation expansion investment}
\nomenclature{$\delta I^{{\rm \zeta}}$}{Converter expansion investment}
\nomenclature{$\delta I^{{\rm j}}$}{Storage expansion investment}
\nomenclature{$\Delta P^{{\rm g,max}}$}{Change in generation capacity}
\nomenclature{$\Delta P^{{\zeta,{\rm max}}}$}{Change in converter capacity}
\nomenclature{$\Delta E^{{{\rm j,max}}}$}{Change in storage capacity}

\nomenclature{$\alpha^{{\ell}}$}{Candidate line binary decision variable}
\nomenclature{$\alpha^{{\ell,\rm te}}$}{Inter-zonal candidate line binary decision variable}
\nomenclature{$\alpha^{{\ell,\rm br}}$}{Intra-zonal candidate line binary decision variable}

\nomenclature{$I^{\ell}$}{Candidate line investment}
\nomenclature{$\A^{\rm br}$}{Balancing responsible optimization variables}
\nomenclature{$\A^{\rm te}$}{Transmission expansion optimization variables}
\nomenclature{$\A_{\rm o}$}{Transmission developer optimization variables}
\nomenclature{$\A_{\rm w}$}{OWPP developer optimization variables}
\nomenclature{$\A_{\rm j}$}{Storage developer optimization variables}

\nomenclature{$E^{{\rm j,max}}$}{Maximum capacity of candidate storage}
\nomenclature{$E^{{\rm j}}$}{Capacity of candidate storage}
\nomenclature{$\widehat{E^{\rm j, max}}$}{Maximum expansion capacity of candidate storage}

\nomenclature{$\pi_s$}{Probability of scenario $s$}
\nomenclature{$\Psi^{\rm g}$}{RES generator time series}

\nomenclature{$\Psi^{\rm u}$}{Demand time series}

\nomenclature{$\eta^{\rm j,inj}$}{Discharge efficiency of storage}
\nomenclature{$\eta^{\rm j,abs}$}{Charge efficiency of storage}
\nomenclature{$\xi^{\rm j,c}$}{Maximum charge rate}
\nomenclature{$\xi^{\rm j,d}$}{Maximum discharge rate}
\nomenclature{$\gamma^{\rm j}$}{Self-discharge rate}

\nomenclature{$T$}{Hours per simulation year}
\nomenclature{$b$}{Transmission line susceptance}
\nomenclature{$\tau$}{Transformer ratio}
\nomenclature{$\theta$}{Voltage angle}
\nomenclature{$\Cand \theta$}{Candidate voltage angle}
\nomenclature{$\theta^{\rm min}$}{Minimum voltage angle}
\nomenclature{$\theta^{\rm max}$}{Maximum voltage angle}
\nomenclature{$\Delta \theta^{\rm max}$}{Maximum voltage angle difference}
\nomenclature{$\Z$}{Set of market zones}
\nomenclature{$L^{\zeta}$}{Converter loss factor}

\nomenclature{$\bm{\Cand P}^{\rm g}$}{Set of candidate generator output variables.}
\nomenclature{$\Delta\bm{\Cand P}^{\rm g,max}$}{Set of change in max generator capacity variables.}
\nomenclature{$\bm\theta$}{Set of voltage angle variables.}
\nomenclature{$\bm\alpha^{\ell}$}{Set of candidate transmission line variables.}
\nomenclature{$\Delta\bm P^{\zeta,\rm max}$}{Set of change in max converter capacity variables.}
\nomenclature{$\bm P^{\rm j,inj}$}{Set of storage injection power variables.}
\nomenclature{$\bm P^{\rm j,abs}$}{Set of storage absorption power variables.}
\nomenclature{$\Delta\bm E^{\rm j,max}$}{Set of change in max storage capacity variables.}
\nomenclature{$\bm{\Exist P}^{\rm g}$}{Set of existing generator output variables.}
\nomenclature{$\bm P^{\rm u}$}{Set of instantaneous demand variables.}

%% file: TeX/abstract.tex
This work examines the Generation and Transmission Expansion (GATE) planning problem of offshore grids under different market clearing mechanisms: a Home Market Design (HMD), a zonal cleared Offshore Bidding Zone (zOBZ) and a nodal cleared Offshore Bidding Zone (nOBZ). It aims at answering two questions.
\begin{enumerate}
    \item Is knowing the market structure a priori necessary for effective generation and transmission expansion planning?
    \item Which market mechanism results in the highest overall social welfare?
\end{enumerate}
To this end a multi-period, stochastic GATE planning formulation is developed for both nodal and zonal market designs. The approach considers the costs and benefits among stake-holders of Hybrid Offshore Assets (HOA) as well as gross consumer surplus (GCS). The methodology is demonstrated on a North Sea test grid based on projects from the European Network of Transmission System Operators' (ENTSO-E) Ten Year Network Development Plan (TYNDP). An upper bound on potential social welfare in zonal market designs is calculated and it is concluded that from a generation and transmission perspective, planning under the assumption of an nOBZ results in the best risk adjusted return.

%% file: TeX/introduction.tex
\vspace{-1mm}\subsection{Motivation}
A perfect storm has descended on the European energy landscape. The ever pressing issues of climate change have converged with an urgent need to reduce the dependence on Russian gas in light of the recent invasion of Ukraine. In response to the invasion, the European Commission recently presented the 300\,B\texteuro{} REpowerEU plan to rapidly scale-up \glspl{RES} and network electrification. The plan builds on the already ambitious targets under the Fit for 55 plan, increasing renewable generation targets to 1236 GW from 1067 GW by 2030 \cite{ec2022repower}.

Offshore wind in the North Sea is crucial in meeting these targets. The North Sea governments of Belgium, Denmark, the Netherlands and Germany have pledged to increase the installed capacity of North Sea offshore wind farms to 65\,GW by 2030 and to 150\,GW by 2050 \cite{jorgensen2022declaration}. This is a substantial step towards the EU wide goal of 240 to 450\,GW of offshore wind by 2050, which is needed to meet the climate targets agreed upon under The Paris Agreement \cite{paris,osullivan2021offshore}.

In addition to expanding offshore wind, investments in transmission infrastructure are required. To this end, the \gls{ENTSO-E} releases a \gls{TYNDP} every two years to identify essential infrastructure investments \cite{entsoe}. To date, 43 offshore transmission projects, including interconnectors, \glspl{HOA} and \gls{OWPP} connections, totalling 65.6\,GW of capacity, are set to be commissioned by 2035. The number of projects is set to increase further as according to EU regulation 2022/869, article 14, by 2024 the TYNDP must include a high level infrastructure investment plan for each of the sea-basins under development \cite{ec2022regulations2}.

\vspace{-1mm}\subsection{Background}
\subsubsection*{Long term planning}
Much research into regulatory, technological and economic aspects of a North Sea grid has been performed \cite{vanhertem2010multi,dedecker2013review,vrana2011technical}. There is a consensus that such a grid would be a meshed \gls{HVDC} grid. The economic and technical advantages of choosing HVDC are summarized in \cite{curis2016deliverable} while some of the technical challenges can be found in \cite{pierri2017challenges}.

The \gls{GATE} planning problem aims at determining the least cost power system design and in its complete form is an \gls{MINLP} \cite{conejo2016generation}. Due to the difficulty of solving such a problem, the reactive power component of the non-linear power flow equations is often ignored and a linear ``DC'' power flow \cite{stott2009dc} or convex relaxation such as  \cite{taylor2012convex,ergun2019optimal,dave2021relaxations} assumed.

In its most common form the problem takes a central planner's perspective as in \cite{baringo2012transmission,zhou2011designing}. Another common way to formulate it is as an equilibrium model such as \cite{chuang2001game,murphy2005generation}. The problem can be formulated as static (single time step) \cite{wang2009strategic,chuang2001game} or dynamic (multi-step) problem \cite{baringo2012wind,botterud2005optimal}. Uncertainty is often handled using stochastic programming as in \cite{lopez2007generation,kazempour2011strategic,kazempour2012strategic} or robust optimization as in \cite{zhang2018robust,jabr2013robust,chen2014robust}. When analyzing energy markets with zonal market clearing, multi-level programming such as in \cite{grimm2016transmission,garces2009bilevel} has been used.

\subsubsection*{Energy Markets}
In liberalized energy markets both nodal and zonal market structures can be found. Examples of the former can be found in North America while in Europe a zonal based system is the norm. For further details on market design, we refer the readers to \cite{entsoe2020,schonheit2021toward}.

A topic currently under debate is the structure of the energy market for an offshore grid and in particular \glspl{HOA} \cite{entsoe2020}. \glspl{HOA} yield benefits for  both \glspl{OWPP} and grid expansion in unison, combining transmission, generation and storage into a single asset. For individual stakeholders, however, there is high uncertainty in expected profitability based on market design and regulation.
Currently, there are three market designs under serious consideration, the \gls{HMD}, the \gls{zOBZ} and the \gls{nOBZ} \cite{ec2020market}.
In an \gls{HMD}, \glspl{HOA} are considered part of the energy market of the \gls{EEZ} within which they are situated. In a \gls{zOBZ}, multiple or all \glspl{HOA} are grouped into a common offshore market zone with a single market price.
In an \gls{nOBZ} all \glspl{HOA} are in independent market zones, permitting fully localized energy pricing based on inter-nodal congestion.

\subsubsection*{Contributions and paper structure}
In this work an agent based, multi-level, multi-period, stochastic, \gls{MILP} is developed. The model builds on the one presented in \cite{ergun2021probabilistic} and leverages the PowerModels(ACDC).jl packages in the Julia programming language to implement the power flow equations \cite{ergun2019optimal,bezanson2017julia,dave2019tnep}. The Gurobi solver version 0.9.14 \cite{gurobi2020gurobi} is used for the MILP.

The main contributions of this work are:
\begin{itemize}
	\item A modelling formulation for different market structures within the \gls{GATE} planning problem.
	\item A measurement of the upper bound on social welfare when considering a zonal market design.
	\item A cost-benefit analysis of a North Sea test case comparing an \gls{HMD}, an \gls{nOBZ} and a \gls{zOBZ}.
\end{itemize}
The structure of the paper is as follows. In the next section a brief discussion on nodal and zonal markets is presented. Following this, the modelling methodology is described. This begins with the nodal market model in \S\ref{sect:nodal_pricing_model} and is followed in \S\ref{sect:zonal_pricing_model} by the zonal market model. In \S\ref{sect:test_cases} the North Sea test grid is described along with the modelling assumptions. In \S\ref{sect:results} the results are presented. Finally, in \S\ref{sect:conclusions}, conclusions and recommendations based on the modelling results round out the paper.

%% file: TeX/nodalVSzonal.tex
In the interest of transparency, the authors of this study declare a pre-existing bias towards nodal pricing due to the price signals in regards to network inefficiencies such as congestion and under supply. These signals can be suppressed in a zonal system and are of very high value, especially at a time of such anticipated network expansion. It is opinioned that localized pricing should be the default and any deviation should be done with caution and only when supported by strong evidence.

Despite the known benefits of nodal pricing, it is possible to find examples where zonal pricing results in an arguably better outcome for consumers. In studying such examples we hope to meaningfully contribute to the ongoing debate regarding North Sea market structure. To illustrate this point we present a simplified market clearing example involving a pivotal supplier. The assumed market structure is a day-ahead pay-as-cleared (uniform pricing method) market \cite{moser2017simulating} with imbalances settled via an idealized \gls{RDCC} \cite{hirth2018market}. Market participants are assumed to bid truthfully at their marginal price of production.

\begin{figure}[!ht]
	\centering
	\begin{minipage}{25mm}
	\includegraphics[width=25mm]{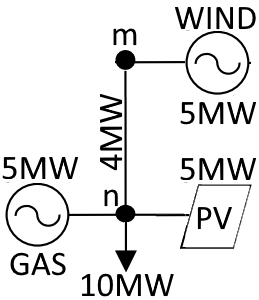}
	\end{minipage}
	\begin{minipage}{62mm}
	\begin{tabular}{lrrrr}
		&\makecell[c]{$P^{\rm g}_{\text{\sc wind}}$\\$[$MW$]$}&\makecell[c]{$P^{\rm g}_{\text{\sc pv}}$\\$[$MW$]$}&\makecell[c]{$P^{\rm g}_{\text{\sc gas}}$\\$[$MW$]$}&\makecell[c]{Cost\\$[$\texteuro{}$]$}\\\hline
		\rowcolor{light-gray}\multicolumn{5}{c}{Nodal}\\
		ID&4.0&5.0&1.0&640\\
		\rowcolor{light-gray}\multicolumn{5}{c}{Zonal}\\
		ID&5.0&5.0&0.0&100\\
		RD&-1.0&0.0&1.0&100\\
		\multicolumn{4}{r}{Zonal total: }&200\\\hline
		\multicolumn{5}{c}{ID: initial dispatch~ RD: re-dispatch}\\
	\end{tabular}
	\end{minipage}
	\caption{Single line diagram of simple market clearing topology (left). Dispatch and re-dispatch amounts and costs in nodal and zonal market clearing mechanisms (right).}
	\label{fig:sld_redispatch}
\end{figure}

In the pivotal supplier scenario, we have a topology similar to that of fig. \ref{fig:sld_redispatch}. At node $m$, 5\,MW of wind generation is present. At node $n$, 5\,MW of PV, 5\,MW of thermal generation and 10\,MW of demand are present. A transmission line with a maximum capacity of 4\,MW connects nodes $m$ and $n$. Assumed marginal generation costs for \glspl{RES} and thermal generation are 10\,\texteuro{}/MWh and 100\,\texteuro/MWh respectively.

In the nodal market clearing model the 4\,MW capacity limit between node $m$ and $n$ is considered from the start, resulting in an optimal dispatch of 4\,MW of wind at node $m$, 5\,MW of PV at node $n$ and 1\,MW of thermal at node $n$. The resulting clearing prices are 10\texteuro{}/MWh at node $m$ and 100\texteuro{}/MWh at node $n$. The total cost of supplying the load is 640\texteuro{}.

Contrasting this with the zonal clearing model, the line congestion between $m$ and $n$ is initially ignored resulting in an optimal dispatch of 5\,MW of wind at node $m$ and 5\,MW of PV at node $n$ with a zonal clearing price of 10\,\texteuro{}/MWh. The cost prior to re-dispatch is therefore 100\,\texteuro{}. As the capacity constraint from node $m$ to $n$ is violated, however, the balancing authority directs the down regulation of wind to 4\,MW at node $m$ and up regulation of the thermal plant at node $n$ to 1~MW. The balancing responsible pays a total re-dispatch cost of 100\,\texteuro{} to the thermal plant and the total cost to supply the load is 200\,\texteuro{} (assuming no avoided variable costs for the \gls{OWPP}).

This is arguably a more desirable result for consumers. Of course, in the case of the pivotal supplier it can be correctly argued that the efficiency of the market is creating a high clearing price to signal that either additional transmission from $m$ to $n$ or additional generation at node $n$ is needed. However, in a grid with high penetration of highly fluctuating sources, it may be an unusually low wind or solar irradiation day that transforms a certain generator into a pivotal supplier. The question as to whether an investment to increase transmission or generation is warranted is more complicated, as it depends on the frequency with which the pivotal supplier negatively impacts energy prices.

%% file: TeX/planning.tex
Special notation:
\begin{itemize}
    \item $x:n$ ($x$ located at node $n$).
    \item $x:mn$ ($x$ located on directed edge $mn$).
    \item $x:\mnm$ ($x$ located on undirected edge $mn$).
\end{itemize}
\vspace{-1mm}\subsection{Nodal Market Model}\label{sect:nodal_pricing_model}
\subsubsection{Objective Function}
An existing network operator ($\A_{\rm e}$) and a set of developers of \glspl{OWPP} ($\A_{\rm w}$), offshore transmission ($\A_{\rm o}$) and storage ($\A_{\rm j}$) have costs and benefits associated with their operation and development as defined in \eqref{eq:wind_objective} to \eqref{eqn:network_objective}. 
The costs and benefits are divided into two distinct parts: hourly operational costs and benefits and yearly strategic investments. The \gls{NPV} equivalents of hourly and yearly revenue and expenditure streams are determined by scalars $f_{y}^{\text{\sc h}}$ and $f_{y}^{\text{\sc y}}$ respectively.
A discount rate of 4\% is assumed.

An \gls{OWPP} developer ($\A_{\rm w}$) can make a strategic yearly investment to expand the capacity of an \gls{OWPP} ($\Delta P_{\Cand g}^{\text{\rm max}}$) as in \eqref{eqn:wind_investment}. Hourly benefits can then be accrued by selling the energy generated on the spot market at a price $\lambda_{n}$. The marginal cost of production is assumed to be zero.
\begin{gather}
	\mathcal{U}^{\rm w}_{y,s}=-f_{y}^{\text{\sc y}}\Big[\eqref{eqn:wind_investment}\Big]
\label{eq:wind_objective}
\end{gather}
\addtocounter{equation}{-1}%
\vspace*{-\baselineskip}%
\begin{subequations}
\begin{gather}
	\sum_{n\in\N^\ac}\sum_{\Cand g\in\Cand\Sgn} \delta I^{\rm g}_{\Cand g:n,y}\cdot \Delta P_{\Cand g:n,y}^{\rm g,max}\label{eqn:wind_investment}
\end{gather}
\end{subequations}
An offshore transmission developer ($\A_{\rm o}$) can make a strategic yearly investment to build new transmission lines ($\alpha^{\ell}_{\Cand l}$) as in \eqref{eqn:tl_invest} and/or expand \gls{HVDC} converter capacity ($\Delta P_{ne}^{\zeta,\rm max}$) as in \eqref{eqn:converter_invest}. Hourly benefits are then accrued through spatial arbitrage of price differentials ($\lambda_{m}-\lambda_{n}$) located in different energy markets. This is also known as congestion rent.%
\begin{gather}
	\mathcal{U}^{\rm o}_{y,s}
{}=
	-f_{y}^{\text{\sc y}}\Big[\eqref{eqn:tl_invest}+\eqref{eqn:converter_invest}\Big]
\label{eqn:tl_objective}
\end{gather}
\addtocounter{equation}{-1}%
\vspace*{-\baselineskip}%
\begin{subequations}
\begin{gather}
	\sum_{\substack{\mnm\subseteq\N\\mn\in\E}}\sum_{\Cand l\in\Cand\Slmnm}\alpha^{\ell}_{\Cand l:\mnm,y}\cdot I^\ell_{\Cand l:\mnm,y}\label{eqn:tl_invest}
\\
	\sum_{ne\in \Eacdc} \delta I^{\zeta}_{ne,y}\cdot \Delta P_{ne,y}^{\zeta,\rm max}\label{eqn:converter_invest}
\end{gather}
\end{subequations}
A storage developer ($\A_{\rm j}$) can make a strategic yearly investment to expand storage capacity ($\Delta E_{j}^{\rm j,max}$) as in  \eqref{eqn:storage_invest}. Hourly benefits are then accrued through temporal arbitrage of price differentials ($\lambda_{n,t}-\lambda_{n,t+\Delta t}$). The marginal cost of charging and discharging is assumed to be zero.
\begin{gather}
	\mathcal{U}^{\rm j}_{y,s}=-f_{y}^{\text{\sc y}}\Big[\eqref{eqn:storage_invest}\Big]
\label{eqn:storage_objective}
\end{gather}
\addtocounter{equation}{-1}%
\vspace*{-\baselineskip}%
\begin{subequations}
\begin{gather}
	\sum_{n\in\N}\sum_{j\in\Sjn} \delta I_{j:n,y}^{\rm j}\cdot \Delta E_{j:n,y}^{\rm j,max}\label{eqn:storage_invest}
\end{gather}
\end{subequations}
The existing network operator has hourly costs and benefits associated with existing generation \eqref{eqn:generation_profit} and consumption \eqref{eqn:consumption_profit}. Existing generators accrue hourly benefits through the sale of energy ($P^{\rm g}_{\Exist g}$) on the spot market at a price ($\lambda_{n}$) higher than their marginal production cost ($C^{\rm g}_{\Exist g}$). Consumers benefit when the price of energy $\lambda_{n}$ is lower than the consumer's bid price ($C^{\rm u}_{u}$) resulting in a surplus.
\begin{gather}
	\mathcal{U}^{\rm e}_{y,s}
{}=
	f_{y}^{\text{\sc h}}\sum_{t\in\St}\Big[\eqref{eqn:generation_profit}+\eqref{eqn:consumption_profit}\Big]
\label{eqn:network_objective}
\end{gather}
\addtocounter{equation}{-1}%
\begin{subequations}
\begin{gather}
	\sum_{n\in\N^\ac}\sum_{\Exist g\in\Exist \Sgn}(-C^{\rm g}_{\Exist g:n,t,y,s})\cdot P^{\rm g}_{\Exist g:n,t,y,s}\label{eqn:generation_profit}
\\
	\sum_{n\in\N^\ac}\sum_{u\in\Sun}C^{\rm u}_{u:n,t,y,s}\cdot P^{\rm u}_{u:n,t,y,s}\label{eqn:consumption_profit}
\end{gather}
\end{subequations}
Combining \eqref{eq:wind_objective}--\eqref{eqn:network_objective} gives the global objective executed by an all knowing centralized authority to maximize the social welfare of the system ($\mathcal{U}$), as in \eqref{eqn:objective_function}. Social welfare is therefore defined as the sum over all scenarios $\Ss$ of \gls{GCS} and net developer benefits. The final distribution of developer benefits and the \gls{GCS} is calculated based on the resultant market clearing prices $\bm \lambda_n$. The uncertainty of long term planning is captured by the probability $\pi_s$ of occurrence of a given scenario. Multi-period planning is performed over the lifetime considering the years defined in set $\Sy$.
\begin{equation}
	\max_{\A_{\rm w},\A_{\rm o},\A_{\rm j},\A_{\rm e}}
	\mathcal{U}:=\sum_{s\in \Ss}{\pi_{s}}\sum_{y\in\Sy}\mathcal{U}_{y,s}^{\rm w}+\mathcal{U}_{y,s}^{\rm o}+\mathcal{U}_{y,s}^{\rm j}+\mathcal{U}^{\rm e}_{y,s}
\label{eqn:objective_function}
\end{equation}
where
\begin{gather*}
	\A_{\rm w}=\bigl(
		\bm{\Cand P}^{\rm g},\,
		\Delta\bm{\Cand P}^{\rm g,max}
	\bigr),~~
\A_{\rm o}=\bigl(
        \bm\theta,\,
		\bm\alpha^{\ell},\,
		\Delta\bm P^{\zeta,\rm max}
	\bigr),\\
	\A_{\rm j}=\bigl(
		\bm P^{\rm j,inj},\,
		\bm P^{\rm j,abs},\,
		\Delta\bm E^{\rm j,max}
	\bigr),~~
	\A_{\rm e}=\bigl(
		\bm{\Exist P}^{\rm g},\,
		\bm P^{\rm u}
	\bigr)
\end{gather*}

\subsubsection{Constraints}
Generation consisting of both \glspl{RES} and conventional generation must remain within capacity limits. This is ensured by constraint \eqref{eqn:gen_capacity_limit}. For \glspl{RES}, parameter $\Psi^{\rm g}_{g}$ is the per-unit \gls{RES} generation time series and for conventional generators it is equal to one.
\begin{equation}
\label{eqn:gen_capacity_limit}
	\begin{array}{@{}c@{}}
		0\leq P^{\rm g}_{g:n,t,y,s}\leq \Psi^{\rm g}_{g:n,t,y,s}\cdot P^{\rm g,max}_{g:n,y}
	\\[5pt]
		 n\in\N^\ac,~~
		 g \in\Sgn,~~
		 t \in \St,~~
		 y \in \Sy,~~
		 s \in \Ss
	\end{array}
\end{equation}
In the particular case that the generator is a candidate \gls{OWPP} under consideration for expansion, $P^{\rm g,max}_{\Cand g}$ is constrained from above by $\widehat{P^{\rm g,max}_{\Cand g}}$ and may only increase or remain constant year over year as in:
\begin{equation}
	\begin{array}{c}
		P^{\rm g,max}_{\Cand g:n,y}\leq\widehat{P^{\rm g,max}_{\Cand g:n}},~~
		P^{\rm g,max}_{\Cand g:n,y-{\Delta y}}{}\leq P^{\rm g,max}_{\Cand g:n,y}\\[3pt]
	\Cand g \in\Cand\Sgn,~~
	 n\in\N^\ac,~~
	 y \in \Sy,
	\end{array}
	\label{eqn:owpp_expansion_constraint}
\end{equation}
where $\Delta y$ is the number of years between modelling years. $P^{\rm g,max}_{\Cand g:n,y-{\Delta y}}$ in the first year is assumed to be zero.
Demand is set via time series $\Psi^{\rm u}_{u}$ as in \eqref{eqn:demand_capacity_limit}. A high cost for load shedding ensures it is only a last resort.
\begin{equation}
\label{eqn:demand_capacity_limit}
\begin{array}{@{}c@{}}
	0\leq P^{\rm u}_{u:n,t,y,s}\leq \Psi^{\rm u}_{u:n,t,y,s}\\[5pt]
	 n\in\N^\ac,~~
	 u \in \Sun,~~
	 t \in\St,~~
	 y \in \Sy,~~
	 s \in\Ss
	\end{array}
\end{equation}

A storage device has a state of charge $E^{\rm j}_{j}$ at each time step $\Delta t$ defined by the following constraint:
\begin{gather}
	\label{eqn:storage_level}
	\begin{array}{@{}r@{}l@{}}
		E^{\rm j}_{j:n,t,y,s}={}& (1-\gamma^{\rm j}_{j:n})^{\Delta t}E^{\rm j}_{j:n,t-\Delta t,y,s}
	\\[5pt]
		&+\Delta t(\eta_{j:n}^{\rm j,abs}P_{j:n,t,y,s}^{\rm j,abs}-\frac{P_{j:n,t,y,s}^{\rm j,inj}}{\eta_{j:n}^{\rm j,inj}})
	\end{array}
\\\notag
	j\in \Sjn,~~  n \in \N^\ac,~~  t \in \St^*,~~  y \in \Sy,~~  s \in \Ss,
\end{gather}
where \(\St^*\) denotes the set of all time steps except the first one.
Here, $\gamma^{\rm j}_{j}$ is the self discharge rate and $\eta_{j}^{\rm j,abs}$ and $\eta_{j}^{\rm j,inj}$ are the charge and discharge efficiencies respectively.

The current state of charge is constrained between zero and the rating of the device $E_{j}^{\rm j,max}$. The device rating can be expanded up to a maximum of $\widehat{E_{j}^{\rm j,max}}$ but may only increase or remain constant year over year as in:
\begin{equation}
\left.
	\begin{array}{@{}r@{}l@{}}
		0\leq E^{\rm j}_{j:n,t,y,s}\leq{}& E_{j:n,y}^{\rm j,max}
	\\[3pt]
		E_{j:n,y-{\Delta y}}^{\rm j,max}\leq E_{j:n,y}^{\rm j,max}\leq{} &\widehat{E_{j:n}^{\rm j,max}}
	\end{array}
\right\}
	\begin{array}{l} n \in \N^\ac, j\in \Sjn\\  t \in \St,  y \in \Sy\\  s \in \Ss.
	\end{array}
\label{eqn:storage_cap_limit}
\end{equation}
In the first year, $E_{j:n,y-{\Delta y}}$ is assumed to be zero. A storage device has a maximum rate at which it can charge and discharge, this is ensured by \eqref{eqn:storage_charge limit}.  $\xi^{\rm j,c}_{j}$ and $\xi^{\rm j,d}_{j}$ are the normalized charge and discharge rates.
\begin{equation}
	\begin{array}{@{}l@{}l@{}l@{}}
		0\leq{} & P_{j:n,t,y,s}^{\rm j,abs} & {}\leq \xi^{\rm j,c}_{j:n}\cdot E_{j:n,y}^{\rm j,max}
	\\[3pt]
		0\leq{} & P_{j:n,t,y,s}^{\rm j,inj} & {}\leq \xi^{\rm j,d}_{j:n}\cdot E_{j:n,y}^{\rm j,max}
	\end{array}
\left\}
	\begin{array}{l} n \in \N^\ac, j\in \Sjn\\  t \in \St, y \in \Sy\\  s \in \Ss
	\end{array}\right.
	\label{eqn:storage_charge limit}
\end{equation}

Constraint \eqref{eqn:storage_boundary} sets the initial and final states of charge in each year to half capacity. The final constraint on storage, to not simultaneously charge and discharge is not explicitly enforced, rather, it is implicitly guaranteed via charge and discharge efficiencies less than one.
\begin{equation}
	\begin{array}{@{}c@{}}
		E^{\rm j}_{j:n,1,y,s}=\frac{E_{j:n,y}^{\rm j,max}}{2}+\eta_{j:n}^{\rm j,abs}P_{j:n,1,y,s}^{\rm j,abs}-\frac{P_{j:n,1,y,s}^{\rm j,inj}}{\eta_{j:n}^{\rm j,inj}}
	\\[3pt]
		E^{\rm j}_{j:n,T,y,s}=\frac{E_{j:n,y}^{\rm j,max}}{2}
	\end{array}
\left\}
	\begin{array}{l@{}} n \in \N^\ac\\ j\in \Sjn\\ y \in \Sy\\  s \in \Ss
	\end{array}\right.
	\label{eqn:storage_boundary}
\end{equation}

The AC and DC network constraints described below are implemented using the PowerModels(ACDC).jl packages \cite{bezanson2017julia,dave2019tnep}.
A bus injection model is used for AC network branches while considering the linear ``DC" power flow approximations as in \eqref{eqn:ac_power_flow}. Transformers are lumped into the branch model via the transformation ratio $\tau$, which is unity when no transformer is required.
\begin{gather}
	\begin{array}{@{}c@{}}\label{eqn:ac_power_flow}
		P^{\ell}_{\Exist l:mn,t,y,s}=\frac{b_{\Exist l:\mnm}}{\tau}[\theta_{m,t,y,s}-\theta_{n,t,y,s}]
	\\[3pt]
		P^{\ell}_{\Cand l:mn,t,y,s}=\frac{b_{\Cand l:\mnm}}{\tau}[\Cand\theta_{\Cand l:mn,t,y,s}-\Cand\theta_{\Cand l:nm,t,y,s}]
	\end{array}\\[5pt]
	\begin{array}{@{}c@{}}\label{eqn:ac_line_capacity_limit}
		|P^{\ell}_{\Exist l:mn,t,y,s}| \leq P_{\Exist l:\mnm}^{\rm \ell,max}\\[3pt]
		|P^{\ell}_{\Cand l:mn,t,y,s}| \leq P_{\Cand l:\mnm}^{\rm \ell,max}\cdot\alpha^{\ell}_{\Cand l:\mnm,y}
 	\end{array}\\[3pt]
	\begin{array}{c}
		 mn \in\E^{\ac},~~
		\Exist l\in\Exist\Slmnm^\ac,~~
		\Cand l\in\Cand\Slmnm^\ac~~\\[3pt]
		 t \in \St,~~
		 y \in \Sy,~~
		 s \in \Ss
	\end{array}\notag
\end{gather}

Power flow through any branch must respect the branch limits as in \eqref{eqn:ac_line_capacity_limit}.
Nodal voltage angle limits and the maximum divergence between connected nodes are constrained by \eqref{eqn:angle_limits}.
\begin{equation}
	\begin{array}{@{}c@{}}
		\theta^{\rm min}\leq\theta_{n,t,y,s}\leq\theta^{\rm max}
	\\[3pt]
		\lvert\theta_{n,t,y,s}-\theta_{m,t,y,s}\rvert\leq\Delta \theta^{\rm max}\\
	\theta^{\rm min}\leq\Cand\theta_{\Cand l:mn,t,y,s}\leq\theta^{\rm max}
	\\[3pt]
	\lvert\Cand\theta_{\Cand l:mn,t,y,s}-\Cand\theta_{\Cand l:nm,t,y,s}\rvert\leq\Delta \theta^{\rm max}
	\\[3pt]
		|\Cand\theta_{\Cand l:mn,t,y,s}-\theta_{m,t,y,s}|\leq (1-\alpha^{\ell}_{\Cand l:\mnm,y})\cdot M
	\end{array}
\left\}
	\begin{array}{l}
		mn \in\E^{\ac}
	\\
		\Cand l\in\Cand\Slmnm^\ac
	\\
		 t \in \St
	\\
		 y \in \Sy
	\\
		 s \in \Ss
	\end{array}\right.
	\label{eqn:angle_limits}
\end{equation}

The final constraint in \eqref{eqn:angle_limits} is only necessary when candidate branches are considered. This constraint leaves candidate line angles unconstrained when the branch is not included, i.e. $\alpha^{\ell}=0$, while enforcing equality with the existing voltage angle when the branch is active, i.e. $\alpha^{\ell}=1$.
The AC network is linked to the DC network via an \gls{HVDC} converter with a maximum AC side capacity of $P^{\zeta,\rm max}$. $P^{\zeta,\rm max}$ is constrained from above by $\widehat{P^{\zeta,\rm max}}$ and can only increase or remain constant year over year as in:
\begin{gather}
	\begin{array}{@{}c@{}}
		|P^{\zeta,\ac}_{ne,t,y,s}|\leq P_{ne,y}^{\zeta,\rm max}
	\\[3pt]
		P_{ne,y-{\Delta y}}^{\zeta,\rm max}\leq P_{ne,y}^{\zeta,\rm max}\leq \widehat{P_{ne}^{\zeta,\rm max}}
	\end{array}
\left\}
	\begin{array}{@{}l@{}} ne\in \Eacdc\\  t \in \St\\ y \in \Sy\\  s \in \Ss.
	\end{array}\right.
	\label{eqn:converter_capacity_limit}
\end{gather}

The AC side power is linked to the DC side power through the non negative converter losses: $L^{\zeta}$. This, and the upper limit on DC side power is set by:
\begin{equation}
\left.
	\begin{array}{@{}c@{}}
		|P^{\zeta,\dc}_{en,t,y,s}|\leq (L^{\zeta}-1)P_{ne,y}^{\zeta,\rm max}
	\\[5pt]
		P_{ne,t,y,s}^{\zeta,\rm loss}=L^{\zeta} P^{\zeta,\ac}_{ne,t,y,s}\geq 0
	\\[5pt]
		P^{\zeta,\ac}_{ne,t,y,s}+P^{\zeta,\dc}_{en,t,y,s}=P_{ne,t,y,s}^{\zeta,\rm loss}
	\end{array}
\right\}
	\begin{array}{@{}l@{}} ne\in \Eacdc\\  t \in \St\\ y \in \Sy\\  s \in \Ss.
	\end{array}
	\label{eqn:converter_cd2ac_link}
\end{equation}

In the DC network, linearized power flow reduces to a network flow model \eqref{eqn:dc_network_flow}. DC side power flow through transmission lines must remain within limits as in \eqref{eqn:dc_line_capacity_limit}.
\begin{gather}
	\begin{array}{@{}c@{}}\label{eqn:dc_network_flow}
	P_{\Exist l:ef,t,y,s}^{\ell}=-P_{\Exist l:fe,t,y,s}^{\ell}\\
	P_{\Cand l:ef,t,y,s}^{\ell}=-P_{\Cand l:fe,t,y,s}^{\ell}
\end{array}\\[5pt]
\begin{array}{@{}c@{}}\label{eqn:dc_line_capacity_limit}
		|P_{\Exist l:ef,t,y,s}^{\ell}|\leq P_{\Exist l:\efe}^{\ell, \rm max}
	\\[3pt]
		|P_{\Cand l:ef,t,y,s}^{\ell}|\leq P_{\Cand l:\efe}^{\ell, \rm max}\cdot\alpha^{\ell}_{\Cand l:\efe,y}
	\end{array}\\[5pt]
\begin{array}{@{}c@{}}
		 ef \in\mathcal E^{\dc},~~
		\Exist l\in\Exist\Slefe^\dc,~~
		\Cand l\in\Cand\Slefe^\dc\\
		 t \in \St,~~
		 y \in \Sy,~~
		 s \in \Ss
	\end{array}\notag
\end{gather}

Finally, Kirchhoff's current law must be satisfied for both AC and DC grids. On the AC side the nodal power balance is given by \eqref{eqn:ac_nodal_equation}. The AC nodal balance equation is a complicating constraint that links the optimization variables. The dual variable of the constraint is $\lambda_{m}$, the marginal price of energy.
\begin{equation}
\left.
	\begin{array}{@{}r<{{}}@{} >{\displaystyle}l@{}>{\displaystyle}l@{}}
	&
		\sum_{e\in \N_m^\dc}P_{me,t,y,s}^{\zeta, \ac}
		{}-{}
		\sum_{n\in\N_m^\ac}\sum_{l\in\Slmnm^\ac}P_{l:mn,t,y,s}^{\ell}
	\\
	+{} &
		\sum_{g\in\Sgn}P_{g:m,t,y,s}^{\rm g}
		{}-{}
		\sum_{u\in\Sun}P_{u:m,t,y,s}^{\rm u}
	\\
	+{} &
		\sum_{j\in\Sjn}P_{j:m,t,y,s}^{\rm j,inj}
		{}-{}
		\sum_{j\in\Sjn}P_{j:m,t,y,s}^{\rm j,abs}=0
	\end{array}
\right\}
	\begin{array}{l@{}} m \in \N^{\ac} \\ t \in \St \\ y \in \Sy \\ s \in \Ss\\(:\lambda_{m,t,y,s})
	\end{array}
	\label{eqn:ac_nodal_equation}
\end{equation}
Here, $\N^{\ac}_m:=\{n\in \N^{\ac}:mn\in\E^\ac\}$ and $\N^{\dc}_m:=\{e\in \N^\dc:me\in\Eacdc\}$ respectively denote the AC and DC neighbors of $m\in\N$. On the DC network side the nodal power balance equation is non complicating:
\begin{gather}
\label{eqn:dc_nodal_equation}
	\sum_{m\in\N_e^\ac}P_{em,t,y,s}^{\zeta, \dc}
	{}+{}
	\sum_{f\in\N_e^\dc}\sum_{l\in\Slefe^\dc}P_{l:ef,t,y,s}^{\ell}
{}={}
	0.
\\
\nonumber
	 e \in  \N^\dc,~~
	 t \in \St,~~
	 y \in \Sy,~~
	 s \in \Ss.
\end{gather}

\vspace{-1mm}\subsection{Zonal Market Model}\label{sect:zonal_pricing_model}
To model a zonal market we start by partitioning the network nodes into a set $\Z$ of disjoint market zones, so that $\N=\bigcup_{z\in\Z}z$.
The edges are also partitioned into inter-zonal edges
$\E^{\rm te}\subseteq\{mn:\,z\in \Z,\,m\in z,\,n\notin z\}$
and intra-zonal edges
$\E^{\rm br}\subseteq\{mn:\,z\in \Z,\,m\in z,\,n\in z\}$.
For ease of notation, the optimization variables for transmission lines along these edges are denoted as ($\theta^{\rm te}$, $\alpha^{\ell, \rm te}$) and ($\theta^{\rm br}$, $\alpha^{\ell, \rm br}$), respectively.
The superscripts refer to two new agents we will introduce, the Transmission Expansion agent ($\A^{\rm te}$) and the Balancing Responsible agent ($\A^{\rm br}$). We re-group the optimization variables into agents $\A^{\rm te}$ and $\A^{\rm br}$ as follows:
\begin{gather*}
\A^{\rm te}=(\A_{\rm w},\A_{\rm j},\A_{\rm e},\A_{\rm o}^{\rm te})
~~\text{and}~~
\A^{\rm br}=(\A_{\rm w},\A_{\rm j},\A_{\rm e},\A_{\rm o}^{\rm br}),
	\label{eqn:br_te_agent_variables}
\end{gather*}
where $\A_{\rm o}^{\rm te}$ differs from $\A_{\rm o}$ in that the transmission lines considered are restricted to inter-zonal connections. $\A_{\rm o}^{\rm br}$ is the complement considering only intra-zonal connections.
Furthermore, we define zonal power balance equations for both the AC and DC networks as in \eqref{eqn:ac_zonal_equation} and \eqref{eqn:dc_zonal_equation}.
\begin{equation}
\left.
	\begin{array}{@{}r<{{}}@{} >{\displaystyle}l@{}>{\displaystyle}l@{}}
	&
		\sum_{m\in z}\bigg(\sum_{e\in \N_m^\dc}P_{me,t,y,s}^{\zeta, \ac}
		{}-{}
		\sum_{g\in\Sgn}P_{g:m,t,y,s}^{\rm g}
	\\
	+{} &
	\sum_{n\in\N_m^{\ac}}\sum_{l\in\Slmnm^{\rm te,\ac}}\hspace{-3.5mm}P_{l:mn,t,y,s}^{\ell}
		{}-{}
		\sum_{u\in\Sun}\hspace{-1.5mm}P_{u:m,t,y,s}^{\rm u}
	\\
	+{} &
		\sum_{j\in\Sjn}P_{j:m,t,y,s}^{\rm j,inj}
		{}-{}
		\sum_{j\in\Sjn}P_{j:m,t,y,s}^{\rm j,abs}\bigg)=0
	\end{array}
\right\}
	\begin{array}{l@{}} z \in\Z,\\ t \in \St, \\ y \in \Sy, \\ s \in \Ss,\\(:\lambda_{z,t,y,s}^{\rm z})
	\end{array}
	\label{eqn:ac_zonal_equation}
\end{equation}

\begin{gather}
\label{eqn:dc_zonal_equation}
		\sum_{e\in z}\bigg(\sum_{m\in\N_e^\ac}P_{em,t,y,s}^{\zeta, \dc}
	{}+{}
	\sum_{f\in\N_e^\dc}\sum_{l\in\Slefe^{\rm te,\dc}}P_{l:ef,t,y,s}^{\ell}\bigg)
{}={}
	0
\\
\nonumber
	 z \in \Z,~~
	 t \in \St,~~
	 y \in \Sy,~~
	 s \in \Ss
\end{gather}

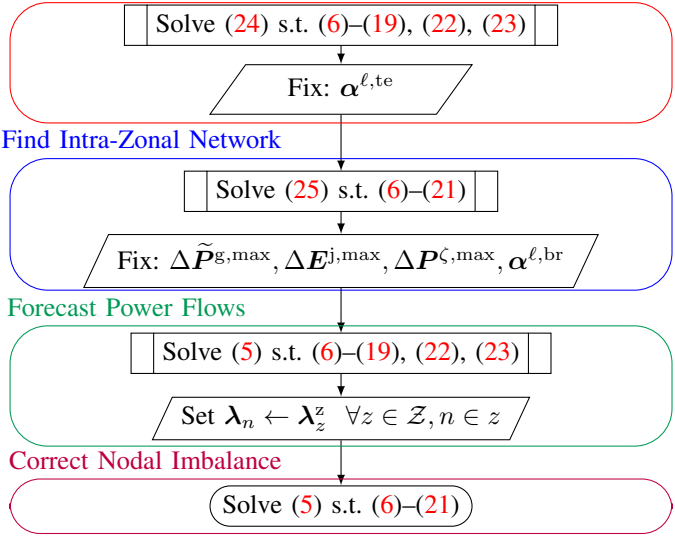
\begin{figure}[hb!]
\centering
	\begin{tikzpicture}

	\node[draw,
		minimum width=3.5cm,
		outer sep=0 ] (block1) at (3,0)  {Solve \eqref{eqn:te_objective} s.t. \eqref{eqn:gen_capacity_limit}--\eqref{eqn:dc_line_capacity_limit}, \eqref{eqn:ac_zonal_equation}, \eqref{eqn:dc_zonal_equation}} ;
	
	\draw (block1.north west) -- ++ (-0.3,0) |- (block1.south west) ;
	
	\draw (block1.north east) -- ++ (0.3,0) |- (block1.south east) ;
		
	\node[draw,
		trapezium,
		trapezium left angle = 65,
		trapezium right angle = 115,
		trapezium stretches,
		below=2.5mm of block1,
		minimum width=3.5cm
	] (block2) {\makecell{Fix: $\bm\alpha^{\ell,\rm te}$}};
	
	\node[draw,
		minimum width=3.5cm,
		below=7.5mm of block2,
		outer sep=0 ] (block3)  {Solve \eqref{eqn:br_objective} s.t. \eqref{eqn:gen_capacity_limit}--\eqref{eqn:dc_nodal_equation}};
	
	\draw (block3.north west) -- ++ (-0.3,0) |- (block3.south west);
	
	\draw (block3.north east) -- ++ (0.3,0) |- (block3.south east) ;
	
	\node[draw,
		trapezium,
		trapezium left angle = 65,
		trapezium right angle = 115,
		trapezium stretches,
		below=3mm of block3,
		minimum width=3.5cm
	] (block4) {\makecell{Fix: $\Delta\bm{\Cand P}^{\rm g,max},\Delta\bm E^{\rm j,max}, \Delta\bm P^{\zeta,\rm max}, \bm\alpha^{\ell,\rm br}$}};

	\node[draw,
		minimum width=3.5cm,
		below=6mm of block4,
		outer sep=0 ] (block5)  {Solve \eqref{eqn:objective_function} s.t. \eqref{eqn:gen_capacity_limit}--\eqref{eqn:dc_line_capacity_limit}, \eqref{eqn:ac_zonal_equation}, \eqref{eqn:dc_zonal_equation}};
	
	\draw (block5.north west) -- ++ (-0.3,0) |- (block5.south west);
	
	\draw (block5.north east) -- ++ (0.3,0) |- (block5.south east) ;
	\node[draw,
		trapezium,
		trapezium left angle = 65,
		trapezium right angle = 115,
		trapezium stretches,
		below=3.1mm of block5,
		minimum width=3.1cm
	] (block6) {Set $\bm\lambda_n\gets\bm\lambda^{\rm z}_z\ \ \forall z\in\Z,n\in z$};
	\node[draw,
		rounded rectangle,
		below=6mm of block6,
		minimum width=3.3cm] (block7) {Solve \eqref{eqn:objective_function} s.t. \eqref{eqn:gen_capacity_limit}--\eqref{eqn:dc_nodal_equation}};
	\draw[-latex]
		(block1) edge (block2)
		(block2) edge (block3)
		(block3) edge (block4)
		(block4) edge (block5)
		(block5) edge (block6)
		(block6) edge (block7);

	\node[draw,
		rounded corners=0.5cm,
		minimum width=8.8cm,
		minimum height=1.6cm,red,label=above:\hspace{-5.3cm} Find Inter-Zonal Network] at (3,-0.5){};

	\node[draw,
		rounded corners=0.5cm,
		minimum width=8.8cm,
		minimum height=1.75cm,blue,label=above:\hspace{-5.3cm}\color{blue}Find Intra-Zonal Network] at (3,-2.65){};

	\node[draw,
		rounded corners=0.5cm,
		minimum width=8.8cm,
		minimum height=1.6cm,ForestGreen,label=above: \hspace{-5.7cm}\color{ForestGreen}Forecast Power Flows] at (3,-4.8){};

	\node[draw,
		rounded corners=0.4cm,
		minimum width=8.8cm,
		minimum height=0.725cm,purple,label=above:\hspace{-5.2cm}\color{purple} Correct Nodal Imbalance] at (3,-6.4){};
		
	\end{tikzpicture}
\caption{Flowchart of solution method for zonal market clearing formulation. {``Solve" refers to the specified equations considering the fixed decision variables found previously.}}
\label{zonal_market_flowchart}
\end{figure}

In a zonal market, \eqref{eqn:ac_zonal_equation} provides the market clearing condition. The dual variable of the constraint is the marginal price of energy at all nodes within zone $z$. As all nodes in a single market zone have a common energy price, the price signals of congestion are suppressed. This further complicates the already difficult problem of \gls{GATE}. To overcome this difficulty a multi-level approach is adopted, whereby expansion planning is achieved via a four step solution method. This solution method is presented in fig. \ref{zonal_market_flowchart}.

\textbf{Step one} is to calculate the inter-zonal network expansion. Agent $\A^{\rm te}$ solves \eqref{eqn:te_objective} (\eqref{eqn:objective_function} with \eqref{eqn:te_tl_invest} substituted for \eqref{eqn:tl_invest}) subject to constraints \eqref{eqn:gen_capacity_limit} through \eqref{eqn:dc_line_capacity_limit} and the zonal power balance constraints \eqref{eqn:ac_zonal_equation} and \eqref{eqn:dc_zonal_equation}. This determines the location and capacity of transmission lines between market zones. The variables $\bm\alpha^{\ell, \rm te}$ are passed to agent $\A^{\rm br}$ to perform intra-zonal transmission and generation expansion.

\begin{gather}
	\max_{\substack{\A^{\rm te}}}{}
	\begin{array}[t]{@{}>{\displaystyle}l@{}}
		\mathcal{U}:=\sum_{s\in \Ss}{\pi_{s}}\sum_{y\in\Sy} f_{y}^{\text{\sc h}}\sum_{t\in \St}\Big[\eqref{eqn:generation_profit}+\eqref{eqn:consumption_profit}\Big]\\-f_{y}^{\text{\sc y}}\Big[\eqref{eqn:wind_investment}+\eqref{eqn:converter_invest}+\eqref{eqn:storage_invest}+\eqref{eqn:te_tl_invest}\Big]
	\end{array}
\label{eqn:te_objective}
\end{gather}
\addtocounter{equation}{-1}%
\vspace*{-\baselineskip}%
\begin{subequations}
\begin{gather}
\sum_{\substack{\mnm\subseteq\N\\mn\in\E^{\rm te}}}\sum_{\Cand l\in\Cand\Slmnm}\alpha^{\ell,\rm te}_{\Cand l:\mnm,y}\cdot I_{\Cand l,y}
\label{eqn:te_tl_invest}
\end{gather}
\end{subequations}

\textbf{Step two} is to calculate intra-zonal expansion. This is performed by agent $\A^{\rm br}$ while respecting the inter-zonal capacity limitations as well as nodal power balance. Hence, objective \eqref{eqn:br_objective} (\eqref{eqn:objective_function} with \eqref{eqn:br_tl_invest} substituted for \eqref{eqn:tl_invest}) is solved subject to nodal power balance \eqref{eqn:ac_nodal_equation} and \eqref{eqn:dc_nodal_equation} as well as the remaining network constraints \eqref{eqn:gen_capacity_limit} through \eqref{eqn:dc_line_capacity_limit}. To achieve nodal balance at the lowest cost, agent $\A^{\rm br}$ has the option to eliminate intra-zonal congestion via the construction of new lines $\alpha^{\ell,\rm br}$, the expansion or reduction of newly added \glspl{OWPP} and/or storage ($\Cand P_{{\Cand g}}^{\rm g, max}, E_{j}^{\rm j, max}$) or the curtailment and/or re-dispatch of network generators ($S_{\rm g}$).

\begin{gather}
	\max_{\substack{\A^{\rm br}}}
	\begin{array}[t]{@{}>{\displaystyle}l@{}}
		\mathcal{U}:=\sum_{s\in \Ss}{\pi_{s}}\sum_{y\in\Sy}f_{y}^{\text{\sc h}}\sum_{t\in \St}\Big[\eqref{eqn:generation_profit}+\eqref{eqn:consumption_profit}\Big]\\-f_{y}^{\text{\sc y}}\Big[\eqref{eqn:wind_investment}+\eqref{eqn:converter_invest}+\eqref{eqn:storage_invest}+\eqref{eqn:br_tl_invest}\Big]
	\end{array}
\label{eqn:br_objective}
\end{gather}
\addtocounter{equation}{-1}%
\vspace*{-.5\baselineskip}%
\begin{subequations}
\begin{gather}
	\sum_{\substack{\mnm\subseteq\N\\mn\in\E^{\rm br}}}\sum_{\Cand l\in \Cand S_{\ell:\mnm}}\alpha^{\ell,\rm br}_{\Cand l:\mnm,y}\cdot I_{\Cand l:\mnm,y}\label{eqn:br_tl_invest}
\end{gather}
\end{subequations}
	
\textbf{Step three} is performed post network expansion. A power auction is held considering only inter-zonal capacity constraints. This step simulates the day ahead spot market and provides the forecast of inter zonal power flows and the zonal market clearing prices based on an optimal dispatch. In the case of congestion free power flows this is the final step. When congestion is present, the additional step of re-dispatch is needed to relieve congestion and balance the network.

\textbf{Step four}: re-dispatch. The final cost of a zonally cleared energy market depends on the re-dispatching mechanism chosen. In the EU both market and regulatory based re-dispatch exists \cite{hirth2018market}. In this work we have chosen an idealized \gls{RDCC} and argue that this represents an upper bound on possible zonal market benefits. Our argument is as follows.

In \gls{RDCC}, an all knowing balancing authority calls on the lowest cost available generation and/or load to up or down regulate during re-balancing actions. Generators required to up regulate are compensated at their marginal price, while those required to down regulate are permitted to keep profits made from the spot market but must return any avoided variable costs (e.g. unused fuel costs). Since participants in \gls{RDCC} are contractually obliged to participate at their marginal rates and we are assuming perfect transparency from generators regarding their marginal cost and available capacity, a market based re-dispatch could only, at best, achieve an identical re-dispatch cost.

Of course, in practice, the assumption of perfect transparency by market participants is unrealistic as this requires the sharing of private information. As such, we are not arguing \gls{RDCC} is better than a market based re-dispatch mechanism. Rather, that in its idealized form it describes the upper bound on efficient re-dispatch.

Together, steps one through four describe the approach used for \gls{GATE} planning in zonal markets. Alone, steps three and four describe an energy auction which can be run on any topology previously determined. For example, a topology determined using the nodal market approach described in section \ref{sect:nodal_pricing_model} can be operated within a zonal market structure.

%% file: TeX/testCase.tex
\vspace{-1mm}\subsection{Domain}
A test grid ($\mathcal{G}$) in the North Sea is modelled. The onshore and offshore nodes considered are displayed in fig. \ref{fig:north_sea_domain} and their coordinates listed in table \ref{tab:test_grid_nodes} of the appendix.

\begin{figure}[h]
    \centering
    \includegraphics[width=52mm]{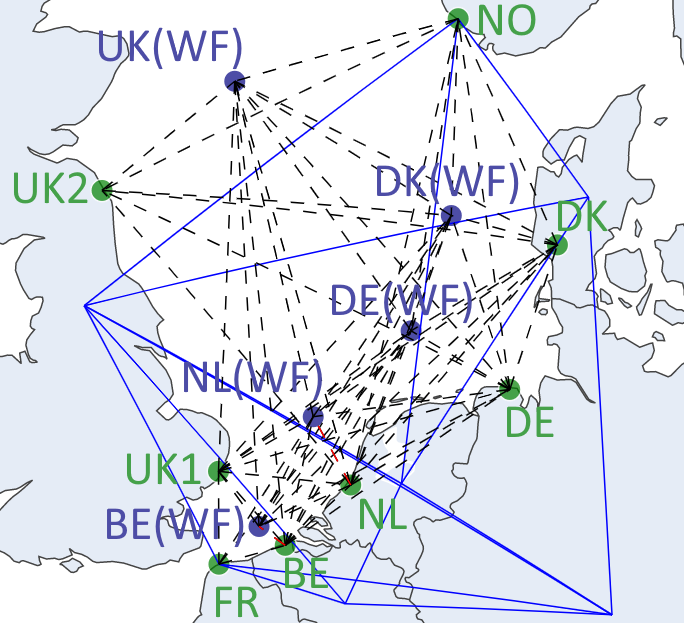}
    \caption{North Sea domain. Lines: NTCs (solid blue), HVDC (dashed black), HVAC (dashed red).}
    \label{fig:north_sea_domain}
\end{figure}
\vspace{-3mm}
\begin{table}[h]
    \centering
    \captionof{table}{Infrastructure costs (excluding cables).}
    \label{tab:cost_of_expansion}
    \begin{tabular}{@{}l|lllll@{}}
         Component&\makecell[c]{OWPPs}&\makecell[c]{Onshore\\converters}&\makecell[c]{Offshore\\converters}&\makecell[c]{Onshore\\storage}&\makecell[c]{Offshore\\storage}\\\hline
         \makecell[c]{Cost}&2100&192.5&577.5&183&275\\
         \multicolumn{6}{l}{*Costs are in \texteuro{}/kW and \texteuro{}/kWh (storage)}
    \end{tabular}
\end{table}
\vspace{-2mm}
\vspace{-1mm}\subsection{Candidate Expansion}
Candidate generation, transmission and storage assets can be expanded at a cost specified in table \ref{tab:cost_of_expansion}. The maximum allowable capacity for \glspl{OWPP} ($\widehat{P^{\rm g, max}}$), HVDC converters ($\widehat{P^{\rm \zeta, max}}$) and storage ($\widehat{E^{\rm j, max}}$) are listed in table \ref{tab:test_grid_nodes} of the appendix. At onshore nodes a dimensioning incident of 3\,GW is assumed, hence onshore converters are limited to 3\,GW while offshore converters can reach sizes of 4\,GW. Storage is assumed to be a four hour duration, lithium ion system. Candidate connections for HVAC and HVDC are shown in fig. \ref{fig:north_sea_domain}. The details of the candidate cables included in each connection as well as their costs are summarized in tables \ref{table:candidate_topology} and \ref{table:ac_cables} of the appendix.
\vspace{-1mm}\subsection{Existing Generation}
The existing energy mix ($\bm{ \Exist P^{\rm g}}$) in each country is sourced from the \gls{ENTSO-E} \gls{TYNDP}, which provides a baseline as well as futures scenarios for the energy mix of European countries. The projected scenarios are for years 2030 and 2040 \cite{entsoedata}, \cite{entsoe}. Further details are provided below. Assumptions for marginal costs of generating sources are listed in table \ref{tab:merit_order} of the appendix.
\vspace{-1mm}\subsection{Onshore grid}
The \glspl{NTC} between the selected countries as provided in the \gls{TYNDP} are modeled as existing transmission lines. They are displayed in fig. \ref{fig:north_sea_domain} and listed in table \ref{tab:ntc} of the Appendix. The onshore \glspl{NTC}
remain static through the simulation years.

\vspace{-1mm}\subsection{Demand}
Hourly demand data ($\bm P^{\rm u}$) is taken from the \gls{TYNDP}. Meeting demand at all times is ideal. When this is not possible, however, load defined as \gls{DSR} can be shed at a cost 119 \texteuro{}/MWh. In the event that further load must be shed, the \gls{VOLL} is 5 k\texteuro{}/MWh. \gls{DSR} does affect market price formation while load shedding does not. During times of extreme energy scarcity such as load shedding events an energy price cap of 180 \texteuro{}/MWh is enforced based on EU regulation 2022/1854 \cite{ec2022regulations}. \gls{GCS} is calculated based on a constant consumer bid price of 150 \texteuro{}/MWh.

\vspace{-1mm}\subsection{Scenarios}
In the \gls{TYNDP}, projections are made for future generation and demand via scenarios that consider different paths towards a net zero 2050. In this study the \gls{NT}, \gls{DG} and \gls{GA} scenarios are included. In brief, \gls{NT} is based on the National Energy Climate Policies, \gls{DG} assumes mass societal adoption of distributed \gls{RES} and \gls{GA} considers a global movement towards the targets of the Paris Agreement. For further details we refer the reader to the \gls{TYNDP} \cite{entsoe}. Pairing historical \gls{RES} generation time series ($\Psi^{\rm g}$) from years 2014 and 2015 results in six scenarios in $\Ss$. Each are considered to have an equally likely probability of occurrence $\pi_s$.

Due to computational requirements each simulation year is clustered into four 24-hour days. The simulation years are 2020, 2030 and 2040. The representative days are found via the $k$-medoids clustering method \cite{kaufman1990partitioning}. The temporal correlation between the various time series is maintained. Ten calendar years per simulation year are considered: 2020-2029 is modelled with 2020 data, 2030-2039 with 2030 data and 2040-2049 with 2040 data.

%% file: TeX/results.tex
\begin{figure*}[!h]
		\begin{minipage}{0.33\textwidth}
		\includegraphics[width=57.75mm]{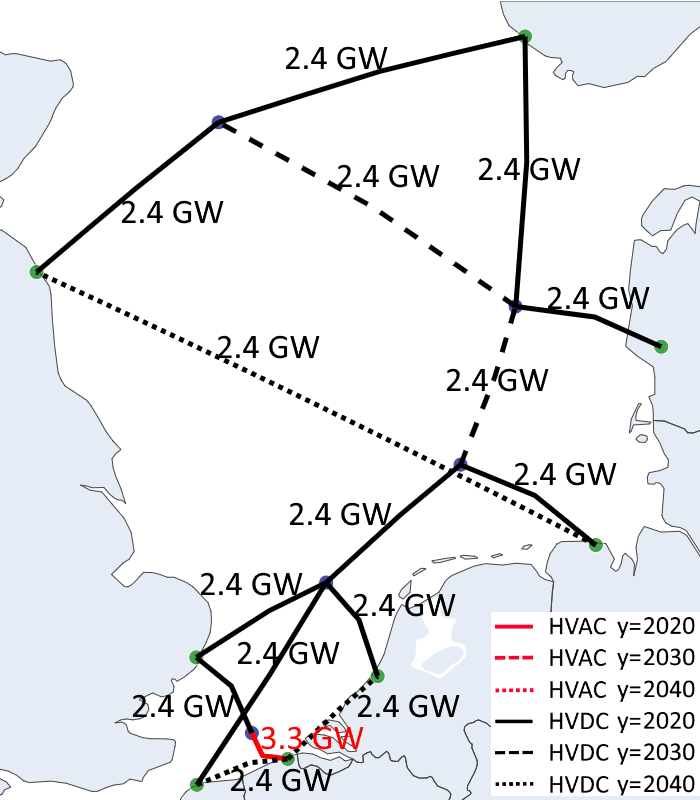}
		\captionof{figure}{nOBZ $\mathcal{G}$ topology.}\label{fig: build_detail_G2_noBZ}
		\end{minipage}
		\begin{minipage}{0.33\textwidth}
		\includegraphics[width=58.5mm]{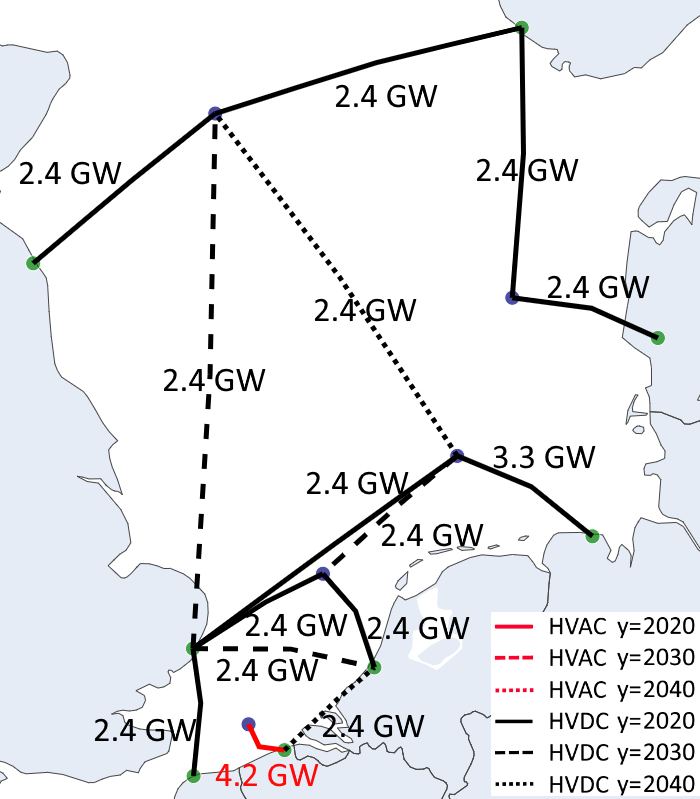}
		\captionof{figure}{HMD $\mathcal{G}$ topology.}\label{fig: build_detail_G2_hmall}
		\end{minipage}
		\begin{minipage}{0.33\textwidth}
		\includegraphics[width=57.75mm]{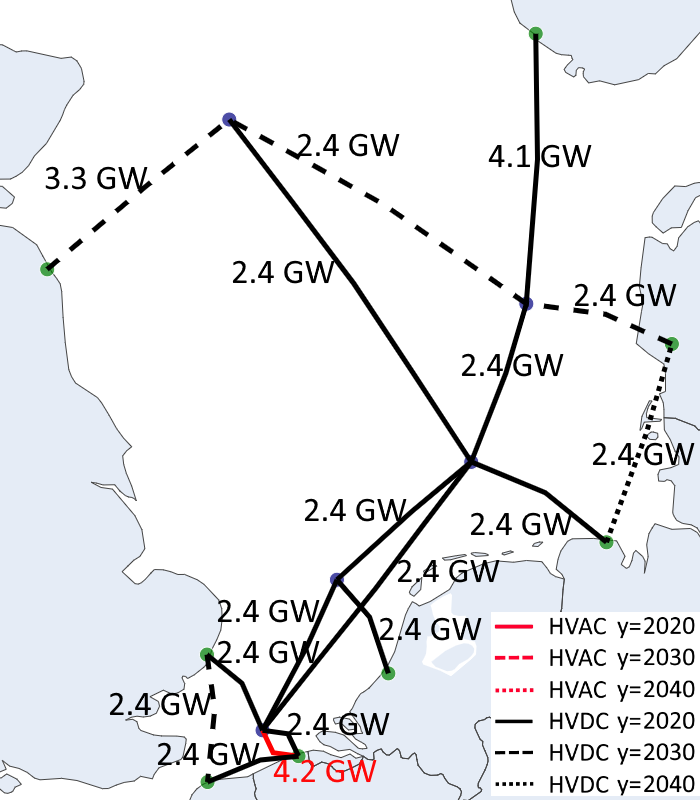}
		\captionof{figure}{zOBZ $\mathcal{G}$ topology.}
		\label{fig: build_detail_G2_zOBZ}
		\end{minipage}
\vspace{-2mm}
\end{figure*}

The presented methodology is applied to $\mathcal{G}$ to compare the effects of an \gls{nOBZ}, \gls{zOBZ} and an \gls{HMD} considering multiple \glspl{OWPP}. The market structure case studies are as follows:
\begin{itemize}
	\item Each \gls{OWPP} is part of its home market zone (\gls{HMD}).
	\item All \glspl{OWPP} form a common offshore market (\gls{zOBZ}).
	\item Each node is its own market zone (\gls{nOBZ}).
\end{itemize}
The resulting topologies are displayed in fig. \ref{fig: build_detail_G2_noBZ} through \ref{fig: build_detail_G2_zOBZ}. The figures present the selected transmission lines and their capacities as well as the build schedule. The same information is provided for the \glspl{OWPP}, HVDC converters and storage in table \ref{tbl: build_detail_G2}. In all topologies, \glspl{HOA} are dominant features. Only in the HMD and only for the closest wind developement region to shore (Belgium) is a radial connection selected.

\begin{table}[!h]
\captionsetup{labelsep=newline,justification=centering,font={sc}}
\caption{Expansion planning schedule of grid $\mathcal{G}$.}
    \label{tbl: build_detail_G2}
    \centering
    \begin{tabular}{@{}m{15mm}@{}D@{}D@{}D@{}|@{}A@{}A@{}A@{}|@{}E@{}E@{}E@{}}
    &\multicolumn{3}{l}{nOBZ}&\multicolumn{3}{l}{HMD}&\multicolumn{3}{l}{zOBZ}\\
       Year & '20 &'30 & '40& '20 &'30 & '40& '20 &'30 & '40\\ \hline
        &\multicolumn{9}{c}{$P_{ne,y}^{\zeta,\rm max}$ [GW]}\\\hline
        UK1 & 2.9 & 3 & 3 & 3 & 3 & 3 & 2.4 & 3 & 3 \\ \hline
        FR & 2.4 & 2.4 & 3 & 2.4 & 2.4 & 2.4 & 2.4 & 3 & 3 \\ \hline
        BE & 0 & 0 & 1.8 & 0 & 0 & 2.4 & 0 & 0.6 & 1.5 \\ \hline
        NL & 2.4 & 2.4 & 3 & 2.4 & 3 & 3 & 2.4 & 2.4 & 2.4 \\ \hline
        DE & 2.4 & 2.4 & 3 & 3 & 3 & 3 & 2.4 & 2.4 & 3 \\ \hline
        DK & 1.6 & 2.4 & 2.4 & 1.7 & 1.9 & 1.9 & 0 & 2.4 & 2.4 \\ \hline
        NO & 3 & 3 & 3 & 3 & 3 & 3 & 3 & 3 & 3 \\ \hline
        UK2 & 1.9 & 2.4 & 3 & 1.7 & 2.4 & 2.4 & 0 & 3 & 3 \\ \hline
        BE(WF) & 0.5 & 0.6 & 0.6 & 0 & 0 & 0 & 0 & 0 & 0 \\ \hline
        DE(WF) & 3.7 & 3.7 & 3.7 & 3.7 & 3.7 & 3.7 & 3.6 & 3.7 & 3.7 \\ \hline
        NL(WF) & 3.8 & 3.8 & 3.8 & 3.8 & 3.8 & 3.8 & 3.8 & 3.8 & 3.8 \\ \hline
        DK(WF) & 3.7 & 3.7 & 3.8 & 3.7 & 3.7 & 3.8 & 3.7 & 3.7 & 3.8 \\ \hline
        UK(WF) & 3.2 & 3.4 & 3.6 & 3.2 & 3.4 & 3.6 & 2.4 & 3.4 & 3.6 \\ \hline
        \multicolumn{10}{l}{\makecell[l]{{\bf OWPPs}: All 4\,GW in 2020 (nOBZ, HMD). \\All 4\,GW in 2020 except UK(WF) is 3.9\,GW in 2020, \\then expanded to 4\,GW in 2030 (zOBZ).\\{\bf Storage}: 1\,GWh of storage is scheduled in Holland in \\2040 (nOBZ, HMD, zOBZ).}}
    \end{tabular}
\vspace{-3mm}
\end{table}

A breakdown of transmission, \gls{OWPP} and storage developer costs and benefits are provided in table \ref{tab: test_grid_2_cost_benefit_summary}. The net benefits, \gls{GCS} and re-dispatch costs of each topology are ranked by social welfare in table \ref{tab: test_grid_2_results_summary}.

\begin{table}[!h]
\captionsetup{labelsep=newline,justification=centering,font={sc}}
\caption{Summary of costs and benefits for $\mathcal{G}$ in B\texteuro.}
\label{tab: test_grid_2_cost_benefit_summary}
	\centering
	\begin{tabular}{@{}lllllll@{}}
		&\multicolumn{2}{l}{Transmission}&\multicolumn{2}{l}{OWPP}&\multicolumn{2}{l}{Storage}\\
		& Cost & Benefits & Cost & Benefits & Cost & Benefits\\\hline
		nOBZ & 21.754 & 64.147 & 42.000 & 134.135 & 0.059 & 0.058 \\
        HMD* & 22.380 & 64.722 & 42.000 & 133.700 & 0.059 & 0.058 \\
        zOBZ* & 22.984 & 70.453 & 41.885 & 125.568 & 0.059 & 0.057 \\
        zOBZ & 22.984 & 57.703 & 41.885 & 135.241 & 0.059 & 0.056 \\
        nOBZ** & 21.754 & 53.452 & 42.000 & 139.866 & 0.059 & 0.057 \\
        nOBZ* & 21.754 & 49.015 & 42.000 & 143.459 & 0.059 & 0.047 \\
        HMD & 22.380 & 48.526 & 42.000 & 144.152 & 0.059 & 0.046 \\ \hline
		\multicolumn{7}{l}{\makecell[l]{HMD*: The HMD topology operating in an nOBZ market.\\zOBZ*: The zOBZ topology operating in an nOBZ market.\\nOBZ*: The nOBZ topology operating in an HMD market.\\nOBZ**: The nOBZ topology operating in a zOBZ market.}}
	\end{tabular}
 \vspace{-3mm}
\end{table}

\begin{table}[h!]
\captionsetup{labelsep=newline,justification=centering,font={sc}}
\caption{Summary of Social Welfare for $\mathcal{G}$ in B\texteuro.}
\label{tab: test_grid_2_results_summary}
	\centering
	\begin{tabular}{llrrrc}
		&\makecell[c]{Net \\Benefit}& GCS &\makecell[c]{Re- \\dispatch}&\makecell[c]{Social \\Welfare}&\makecell[c]{Difference \\$[$\%$]$}\\\hline
		nOBZ & 134.527 & 1920.331 & 0.000 & 2054.858 & - \\
        HMD* & 134.040 & 1920.299 & 0.000 & 2054.340 & -0.03 \\
        zOBZ* & 131.151 & 1922.617 & 0.000 & 2053.767 & -0.05 \\
        zOBZ & 128.073 & 1926.606 & 128.596 & 1926.083 & -6.27 \\
        nOBZ** & 129.561 & 1921.987 & 127.006 & 1924.543 & -6.34 \\
        nOBZ* & 128.707 & 1924.729 & 145.615 & 1907.821 & -7.16 \\
        HMD & 128.284 & 1916.556 & 155.168 & 1889.672 & -8.04 \\ \hline
        \multicolumn{6}{l}{\makecell[l]{%
			*-entries in the first column are as in table \ref{tab: test_grid_2_cost_benefit_summary}.
		}}
		\\
        \multicolumn{6}{l}{\makecell[l]{%
			The last column is the change in social welfare compared to the nOBZ.%
		}}

	\end{tabular}
 \vspace{-3mm}
\end{table}
There is little difference between the social welfare obtained under all variations of the nodal market structure (nOBZ, HMD* and zOBZ*). Three different topologies, all with similar levels of social welfare and a global lower bound (nOBZ), effectively demonstrate the flatness of the solution space and hence the limited value attached to the certificate of optimality. Planners should therefore not be overly concerned about developing a uniquely optimal configuration as the problem's uncertainty dwarfs the difference between a good solution and the best solution.

The benefit of using a nodal based pricing mechanism is clearly demonstrated. All zonal pricing models result in a decrease in social welfare of 6-8\% compared to nodal pricing. The worst performing market structure is the \gls{HMD}. It seems, knowing the market structure a priori is not essential from a planning perspective as each design operates relatively well under a changing market design.

Despite the poor performance of the zonal market models in the zonal markets for which they were intended, the high quality of their nodal market variations suggest merit from a decomposition perspective. The computation times for the approaches are: zOBZ/HMD: $\approx$2.5 hours, the nOBZ: halted after twelve hours with a small optimality gap of 0.04\% remaining. Zonal models scale better than nodal due to the natural decomposition along intra and inter zonal lines.

The three nodal market models may have little variation in social welfare but do have variation in how agent benefits are distributed. For example, nOBZ results in 6.8\% higher benefits for an \gls{OWPP} developer than in zOBZ*. There is no free lunch of course as this is at the expense of the transmission developer which sees a decrease in benefits of 9\%. The ability to adjust the distribution of benefits among stakeholders without sacrificing overall social welfare may prove useful.

All market mechanisms result in a positive return on investment for all agents (fig. \ref{fig: test_grid_2_roi_appendix}). The highest return for transmission and storage developers occurs under an nOBZ  while \gls{OWPP} developers do best in an HMD.
By examining the average energy prices per node in figs. \ref{fig: energy_onshore_prices_G2} and \ref{fig: energy_offshore_prices_G2} we see why. The HMD has the highest average energy prices both offshore and onshore. While this translates to higher profits for \gls{OWPP} developers, it is bad for consumers and lowers the overall social welfare. The average European wide energy price for the HMD is 93.23 \texteuro{}/MWh, for the zOBZ it is 91.21 \texteuro{}/MWh and for the nOBZ it is 90.86 \texteuro{}/MWh.

\begin{figure}[t!]
    \centering
\begin{tikzpicture}
\begin{axis}[
xticklabels={nOBZ, HMD, zOBZ},
xtick=data,
major x tick style=transparent,
ybar,
ymax=11,
ymin=0,
width=40mm,
height=35mm,
bar width = 3.5mm,
x axis line style={opacity=0},
x tick label style={rotate=0, anchor=center},
xticklabel style={yshift=-2mm},
xtick=data,
ymajorgrids=false,
grid style=dotted,
nodes near coords,
scale only axis,
point meta=explicit symbolic,
enlarge x limits = {abs=1},
title style={at={(0.495,0.8)},font=\small},
	yticklabel= {\empty},
	major y tick style=transparent,
	xticklabel style={font=\small},
cycle list={
	fill=red,fill opacity=0.6,nodes near coords style={xshift=6,yshift=-8, black,font=\small,rotate=90}\\
	fill=orange,fill opacity=0.6,nodes near coords style={xshift=6,yshift=-10,rotate=90,black,font=\small}\\
	fill=green,fill opacity=0.6,nodes near coords style={xshift=6,yshift=-10,black,font=\small,rotate=90}\\
},
legend columns=-1,
legend style={at={(1.05,-0.2)},font=\scriptsize},
]

\addplot  coordinates {
(1,7.95)[7.95]
(3,8.20)[8.20]
(5,7.98)[7.98]};

\addplot  coordinates {
(1,7.67)[7.67]
(3,6.61)[6.61]
(5,7.12)[7.12]};
			
\addplot  coordinates {
(1,3.92)[3.92]
(3,3.13)[3.13]
(5,3.82)[3.82]};
	
\legend{OWPP,Transmission,Storage}
\end{axis}
\end{tikzpicture}
\captionof{figure}{Yearly percent return on investment for $\mathcal{G}$.}
\label{fig: test_grid_2_roi_appendix}
\end{figure}
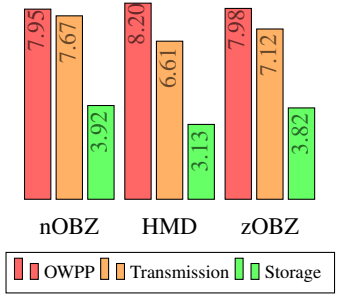

\begin{figure}[t!]
\vspace{-10mm}
\begin{minipage}{88mm}
\centering
\begin{tikzpicture}
\begin{axis}[
xticklabels={UK, FR, BE, NL, DE, DK, NO, All},
xtick=data,
major x tick style=transparent,
ybar,
ymax=112,
ymin=80,
width=85mm,
height=32mm,
bar width = 2.6mm,
x axis line style={opacity=0},
x tick label style={rotate=0, anchor=center},
xticklabel style={yshift=-2mm},
xtick=data,
ymajorgrids=false,
grid style=dotted,
nodes near coords,
scale only axis,
point meta=explicit symbolic,
enlarge x limits = {abs=1},
title style={at={(0.495,0.9)},font=\small},
	yticklabel= {\empty},
	major y tick style=transparent,
	xticklabel style={font=\small},
cycle list={
	fill=red,fill opacity=0.6,nodes near coords style={xshift=6.5,yshift=-13, black,font=\small,rotate=90}\\
	fill=orange,fill opacity=0.6,nodes near coords style={xshift=6.5,yshift=-13,rotate=90,black,font=\small}\\
	fill=green,fill opacity=0.6,nodes near coords style={xshift=6.5,yshift=-13,black,font=\small,rotate=90}\\
},
legend columns=-1,
legend style={at={(0.7,-0.2)},font=\scriptsize}
]

\addplot  coordinates {
(1,92.77)[92.77]
(3,88.54)[88.54]
(5,90.62)[90.62]
(7,91.68)[91.68]
(9,92.58)[92.58]
(11,92.65)[92.65]
(13,106.32)[106.32]
(15,93.59)[93.59]};		
           						
\addplot  coordinates {
(1,92.62)[92.62]
(3,89.06)[89.06]
(5,92.13)[92.13]
(7,92.50)[92.50]
(9,92.32)[92.32]
(11,93.01)[93.01]
(13,105.92)[105.92]
(15,93.94)[93.94]};

\addplot  coordinates {
(1,92.50)[92.50]
(3,88.36)[88.36]
(5,90.05)[90.05]
(7,91.45)[91.45]
(9,92.56)[92.56]
(11,92.39)[92.39]
(13,106.25)[106.25]
(15,93.37)[93.37]};

\legend{nOBZ,HMD,zOBZ}
\end{axis}
\end{tikzpicture}
		\caption{Average onshore energy prices for $\mathcal{G}$ [\texteuro/MWh].}
		\label{fig: energy_onshore_prices_G2}
\end{minipage}
\begin{minipage}{88mm}
\centering
\begin{tikzpicture}
\begin{axis}[
xticklabels={BE(WF), NL(WF), DE(WF), DK(WF), UK(WF), All},
xtick=data,
major x tick style=transparent,
ybar,
ymax=95,
ymin=70,
width=80mm,
height=15mm,
bar width = 2.75mm,
x axis line style={opacity=0},
x tick label style={rotate=0, anchor=center},
xticklabel style={yshift=-2mm},
xtick=data,
ymajorgrids=false,
grid style=dotted,
nodes near coords,
scale only axis,
point meta=explicit symbolic,
enlarge x limits = {abs=1},
title style={at={(0.495,0.9)},font=\small},
	yticklabel= {\empty},
	major y tick style=transparent,
	xticklabel style={font=\small},
cycle list={
	fill=red,fill opacity=0.6,nodes near coords style={xshift=6.5,yshift=-13, black,font=\small,rotate=90}\\
	fill=orange,fill opacity=0.6,nodes near coords style={xshift=6.5,yshift=-13,rotate=90,black,font=\small}\\
	fill=green,fill opacity=0.6,nodes near coords style={xshift=6.5,yshift=-13,black,font=\small,rotate=90}\\
},
legend columns=-1,
legend style={at={(0.715,-0.4)},font=\scriptsize}
]

\addplot  coordinates {
(1,90.40)[90.40]
(3,88.99)[88.99]
(5,85.04)[85.04]
(7,86.95)[86.95]
(9,89.26)[89.26]
(11,88.13)[88.13]};

\addplot  coordinates {
(1,92.13)[92.13]
(3,92.50)[92.50]
(5,92.32)[92.32]
(7,93.01)[93.01]
(9,92.62)[92.62]
(11,92.52)[92.52]};
				
\addplot  coordinates {
(1,89.06)[89.06]
(3,89.06)[89.06]
(5,89.06)[89.06]
(7,89.06)[89.06]
(9,89.06)[89.06]
(11,89.06)[89.06]};
\legend{nOBZ,HMD,zOBZ}
\end{axis}
\end{tikzpicture}
		\caption{Average offshore energy prices for $\mathcal{G}$ [\texteuro/MWh].}
		\label{fig: energy_offshore_prices_G2}
		\end{minipage}
  \vspace{-4mm}
		\end{figure}
\begin{figure}[t!]
\centering
\begin{minipage}{88mm}
\centering
\begin{tikzpicture}
\begin{axis}[
xticklabels={nOBZ, HMD, zOBZ},
xtick=data,
major x tick style=transparent,
ybar,
ymax=11.801,
ymin=0,
width=35mm,
height=25mm,
bar width = 3.5mm,
x axis line style={opacity=0},
x tick label style={rotate=0, anchor=center},
xticklabel style={yshift=-2mm},
xtick=data,
ymajorgrids=false,
grid style=dotted,
nodes near coords,
scale only axis,
point meta=explicit symbolic,
enlarge x limits = {abs=1},
title style={at={(0.495,0.8)},font=\small},
	yticklabel= {\empty},
	major y tick style=transparent,
	xticklabel style={font=\small},
cycle list={
	fill=red,fill opacity=0.6,nodes near coords style={xshift=6,yshift=-11, black,font=\small,rotate=90}\\
	fill=orange,fill opacity=0.6,nodes near coords style={xshift=6,yshift=-10,rotate=90,black,font=\small}\\
},
legend columns=-1,
legend style={at={(1.07,-0.25)},font=\scriptsize},
]

\addplot  coordinates {
(1,3.3)[3.3]
(3,3.35)[3.35]
(5,11.8)[11.8]};

\addplot  coordinates {
(1,3.51)[3.51]
(3,3.51)[3.51]
(5,3.518)[3.518]};
			
\legend{OWPP curtailment, Lost load}
\end{axis}
\end{tikzpicture}
\caption{OWPP curtailment and lost load in TWh. {Potential OWPP production: 2287\,TWh (nOBZ/HMD), 2284\,TWh (zOBZ). Total load: 50366\,TWh.}}
\label{fig: test_grid_2_curtailment_voll}
%


    \centering

    \begin{tikzpicture}
	\centering
	\begin{axis}[
	font=\small,
		ybar stacked,
		ymin=0,
		legend style={at={(0.97,-0.15)},font=\scriptsize},
		legend columns=-1,
		bar width = 10mm,
		width = 70mm,
		height = 50mm,
		enlarge x limits=0.25,
		major x tick style=transparent,
		symbolic x coords={HMD, zOBZ, nOBZ*, nOBZ**},
		xtick=data,
		xticklabel style={font=\footnotesize},
		nodes near coords,
		nodes near coords align={anchor=south},
		x axis line style={opacity=0},
		yticklabel= {\empty},
	major y tick style=transparent,
	]
		\addplot [fill=green,
		nodes near coords align={anchor=south}] coordinates {
	({HMD},9.63)
	({zOBZ},0.00)
	({nOBZ*},0.00)
	({nOBZ**},0.00)};
	
	\addplot [fill=red,nodes near coords align={anchor=south},fill opacity=0.6] coordinates {
	({HMD},56.10)
	({zOBZ},56.57)
	({nOBZ*},54.38)
	({nOBZ**},59.06)};
	
	\addplot [fill=orange] coordinates {
	({HMD},3.79)
	({zOBZ},5.12)
	({nOBZ*},4.24)
	({nOBZ**},5.06)};

	
	\addplot [fill=yellow,nodes near coords align={anchor=south},] coordinates {
	({HMD},26.01)
	({zOBZ},38.09)
	({nOBZ*},36.45)
	({nOBZ**},35.68)};
	
	\addplot [fill=aliceblue,nodes near coords align={anchor=south},] coordinates {
	({HMD},4.47)
	({zOBZ},0.00)
	({nOBZ*},4.76)
	({nOBZ**},0.00)};

	\legend{RES, CCGT, OCGT, REST, VOLL}
	\end{axis}
	\end{tikzpicture}

    \caption{Percent breakdown of re-dispatch costs.}
    \label{fig:redispatch_costs}
    \end{minipage}
    \vspace{-4mm}
    \end{figure}

There is very little difference in wind curtailment between market designs, which amounts to essentially zero. This is shown in fig. \ref{fig: test_grid_2_curtailment_voll}. Although the zOBZ results in about three times the curtailment compared to the nOBZ and the HMD, it is still only about 0.5\% of the total energy production.

The cost to re-dispatch varies substantially between zonal market models. The highest re-dispatch costs are associated with Home market designs and the lowest with zonal OBZs. A breakdown of the re-dispatch costs by generation type is presented in fig. \ref{fig:redispatch_costs}. It is interesting that in the home market designs load shedding makes up about 5\% of the overall re-dispatch cost. It is not that more load is shed in this market structure, as load shedding is fairly constant at about 3.5\,TWh across all market models (fig. \ref{fig: test_grid_2_curtailment_voll}), but rather that the onshore-offshore congestion constraint is binding but not enforced. This is a variation on the pivotal supplier problem discussed in section \ref{sect: pivotal_supplier} with the exception that VOLL does not affect the spot price formulation so the zOBZ and nOBZ** see the energy price cap of 180 \texteuro{}/MWh rather than the 5k\texteuro{}/MWh for VOLL avoiding windfall profits for generators.

It is worth discussing the low level of storage selected in all topologies. This may be due to too high a cost or is possibly a modelling deficiency as type specific generation constraints are not implemented. Ramping times, start up costs and seasonality are completely ignored. To fully capture the benefits of storage some further level of detail may be necessary. For example, Norway has over 24\,GW of pumped storage that is highly seasonal. Modelling this seasonality constraint (among others) might be necessary to effectively access the value of additional storage assets. This investigation is saved for future work.

%% file: TeX/conclusion.tex
	In this work, a generation and transmission expansion planning model for nodal and zonal market designs is developed, with the objective of maximizing a social welfare function consisting of net benefits for the generation, storage and transmission system developers and \gls{GCS}. The developed model is applied to a test case in the North Sea ($\mathcal{G}$) to investigate the impact of market design on the optimal network topology.

From our study of $\mathcal{G}$ we draw the following conclusions. \glspl{HOA} are an important and cost effective feature of offshore transmission topologies.
    The nodal pricing structure results in the highest social welfare. The solution space around the optimal nodal market solution is quite flat and multiple solutions of acceptable quality exist. A zonal planning approach has computational
advantages over a nodal one and may be an effective decomposition strategy for larger problems.
    A topology developed assuming one market structure can operate well under a different market structure.
    The lowest European wide average energy price occurs in the nOBZ and the highest in the HMD.
    The pivotal supplier effect is present in re-dispatch of zonal markets as demonstrated in the HMD.

    From the conducted analysis we conclude that generation and transmission planning should be carried out under the assumption that a nodal offshore bidding zone will be implemented as the obtained topology also performs well under changing market designs. When computational power becomes a constraint, however, a decomposition strategy based on a zonal market design may be used.
More efficient and structure-exploiting optimization strategies, based e.g. on fast consensus ADMM \cite{themelis2022douglas} or Bender's decomposition \cite{baringo2012wind}, are also considered for future extensions.

%% file: TeX/appendix.tex
\begin{table}[h]
\captionsetup{labelsep=newline,justification=centering,font={sc}}
\caption{HVAC and HVDC Candidate transmission lines}
	\centering
	\begin{tabular}{@{}ccc|c|c@{}}
	\hline
\multicolumn{3}{l|}{Routes}&\multicolumn{2}{l}{Candidate Cables ($S_{\ell}$)}\\\hline
		start & end & km & $\mathcal{G}$ \\ \hline
		UK1 & FR & 175 &  DC1-DC3 \\
        UK1 & BE & 188 &  DC1-DC3 \\
        UK1 & NL & 250 &  DC1-DC3 \\
        UK1 & DE & 565 &  DC1-DC3 \\
        UK1 & DK & 754 &  DC1-DC3 \\
        UK1 & BE(WF) & 129 &  DC1-DC3 \\
        UK1 & DE(WF) & 443 &  DC1-DC3 \\
        UK1 & NL(WF) & 204 &  DC1-DC3 \\
        UK1 & DK(WF) & 639 &  DC1-DC3 \\
        UK1 & UK(WF) & 716 &  DC1-DC3 \\
        FR & BE & 130 &  DC1-DC3 \\
        FR & DE(WF) & 565 &  DC1-DC3 \\
        FR & NL(WF) & 328 &  DC1-DC3 \\
        BE & NL & 168 &  DC1-DC3 \\
        BE & DE & 511 &  DC1-DC3 \\
        BE & DK & 752 &  DC1-DC3 \\
        BE & BE(WF) & 61 &  DC1-DC3, AC1-AC3 \\
        BE & DE(WF) & 464 &  DC1-DC3 \\
        BE & NL(WF) & 247 &  DC1-DC3 \\
        BE & DK(WF) & 684 &  DC1-DC3 \\
        BE & UK(WF) & 859 &  DC1-DC3 \\
        NL & DE & 346 &  DC1-DC3 \\
        NL & DK & 586 &  DC1-DC3 \\
        NL & NO & 873 &  DC1-DC3 \\
        NL & UK2 & 713 &  DC1-DC3 \\
        NL & BE(WF) & 189 &  DC1-DC3 \\
        NL & DE(WF) & 308 &  DC1-DC3 \\
        NL & NL(WF) & 146 &  DC1-DC3, AC4, AC5 \\
        NL & DK(WF) & 531 &  DC1-DC3 \\
        NL & UK(WF) & 770 &  DC1-DC3 \\
        DE & DK & 280 &  DC1-DC3 \\
        DE & NO & 679 &  DC1-DC3 \\
        DE & UK2 & 834 &  DC1-DC3 \\
        DE & BE(WF) & 534 &  DC1-DC3 \\
        DE & DE(WF) & 212 &  DC1-DC3 \\
        DE & NL(WF) & 369 &  DC1-DC3 \\
        DE & DK(WF) & 337 &  DC1-DC3 \\
        DE & UK(WF) & 753 &  DC1-DC3 \\
        DK & NO & 444 &  DC1-DC3 \\
        DK & UK2 & 836 & DC1-DC3 \\
        DK & BE(WF) & 761 &  DC1-DC3 \\
        DK & DE(WF) & 311 & DC1-DC3 \\
        DK & NL(WF) & 550 &  DC1-DC3 \\
        DK & DK(WF) & 201 &  DC1-DC3 \\
        DK & UK(WF) & 654 &  DC1-DC3 \\
        NO & UK2 & 711 & DC1-DC3 \\
        NO & DE(WF) & 571 & DC1-DC3 \\
        NO & DK(WF) & 353 & DC1-DC3 \\
        NO & UK(WF) & 414 & DC1-DC3 \\
        UK2 & DK(WF) & 638 & DC1-DC3 \\
        UK2 & UK(WF) & 311 & DC1-DC3 \\
        BE(WF) & DE(WF) & 462 & DC1-DC3 \\
        BE(WF) & NL(WF) & 229 & DC1-DC3 \\
        BE(WF) & DK(WF) & 677 & DC1-DC3 \\
        BE(WF) & UK(WF) & 820 & DC1-DC3 \\
        DE(WF) & NL(WF) & 241 & DC1-DC3 \\
        DE(WF) & DK(WF) & 223 & DC1-DC3 \\
        DE(WF) & UK(WF) & 556 & DC1-DC3 \\
        NL(WF) & DK(WF) & 449 & DC1-DC3 \\
        NL(WF) & UK(WF) & 630 & DC1-DC3 \\
        DK(WF) & UK(WF) & 460 & DC1-DC3 \\\hline
		\multicolumn{5}{l}{\makecell[l]{Note: The specified lengths are 125\% of the Euclidean \\distances to account for obstructions in the shortest path.}}
	\end{tabular}
	\label{table:candidate_topology}
\end{table}
\begin{table}[h]
\footnotesize
\captionof{table}{HVAC(DC) candidate cables \cite{promotion2020cost,abb2006hvdc}.}
		\centering
		\begin{tabular}{llrrr}
		\hline
		&n$\cdot$cm$^{2}$&MVA&km&Cost\\\hline
		AC1&12$\cdot$16&4213&61&1520\,M\texteuro\\
		AC2&11$\cdot$10&3319&61&1065\,M\texteuro\\
		AC3&8$\cdot$10&2414&61&785\,M\texteuro\\
		AC4&11$\cdot$16&3236&146&3345\,M\texteuro\\
	AC5&12$\cdot$6.3&2479&146&2338\,M\texteuro\\
	DC1&4$\cdot$15&4085&-&3.593\,M\texteuro/km\\
		DC2&4$\cdot$10&3288&-&3.194\,M\texteuro/km\\
		DC3&2$\cdot$20&2407&-&2.575\,M\texteuro/km\\\hline
		\multicolumn{5}{l}{\makecell[l]{*Specific HVAC cables are listed since capacity\\ is dependent on length due to reactive power.}}\\
		\end{tabular}
		\label{table:ac_cables}
\end{table}
\begin{table}[h]
\footnotesize
	\captionof{table}{Marginal price of generators \cite{entsoedata}.}
	\centering
	\begin{tabular}{lc}
		Generation Type&\texteuro /MWh\\\hline
		\makecell[l]{PV, Hydro}&18\\
		Onshore wind&25\\
		Offshore wind&59\\
		Other RES&60\\
		Gas CCGT&89\\
		Nuclear&110\\
		DSR&119\\
		\makecell[l]{Gas OCGT, Coal,\\ Pump storage, P2G,\\Other non-RES}&120\\
		Light oil&140\\
		\makecell[l]{Heavy oil, Shale oil}&150\\\hline
	\end{tabular}

	\label{tab:merit_order}
\end{table}
\begin{table}[h]
    \centering
    \begin{tabular}{ccccc}\hline
         UK-FR&UK-BE&UK-NL&UK-DE&UK-DK\\
         4&1&1&1.4&1.4\\
         UK-NO&FR-BE&FR-DE&BE-NL&BE-DE\\
         2.8&4.3&3&2.4&1\\
         NL-DE&NL-DK&NL-NO&DE-DK&DK-NO\\
         5&0.7&0.7&3.5&1.64\\\hline
    \end{tabular}
    \caption{Net transfer capacities between countries in GW \cite{entsoedata}.}
    \label{tab:ntc}
\end{table}
\begin{table}[h]
\captionsetup{labelsep=newline,justification=centering,font={sc}}
\caption{Location of nodes in test grids and maximum capacity of candidate Infrastructure.}
	\label{tab:test_grid_nodes}
	\centering
	\begin{tabular}{llrccc}\hline
		Point&Longitude&Latitude&\makecell[c]{$\widehat{P^{\zeta,\rm max}}$\\$[$GW$]$}&\makecell[c]{$\widehat{P^{\rm g, max}}$\\$[$GW$]$}&\makecell[c]{$\widehat{E^{\rm j, max}}$\\$[$GWh$]$}\\\hline
		UK1&52.21025&1.57374&3.0&-&1.0\\
		FR&50.96332&1.82967&3.0&-&1.0\\
		BE&51.32081&3.20768&3.0&-&1.0\\
		NL&52.22215&4.49556&3.0&-&1.0\\
		DE&53.67043&7.84620&3.0&-&1.0\\
		DK&55.61420&8.72899&3.0&-&1.0\\
		NO&58.43791&6.00292&3.0&-&1.0\\
		UK2&55.68940&-1.91052&3.0&-&1.0\\
		BE(WF)&51.53509&2.59644&4.0&4.0&0.02\\
		NL(WF)&53.08300&3.51802&4.0&4.0&0.02\\
		DE(WF)&54.34610&5.52400&4.0&4.0&0.02\\
		DK(WF)&55.90115&6.22240&4.0&4.0&0.02\\
		UK(WF)&57.33721&0.81425&4.0&4.0&0.02\\\hline
	\end{tabular}
\end{table}